%% file: main.tex
\documentclass[a4paper,USenglish,pdfa,thm-restate]{lipics-v2021}

\pdfoutput=1
\hideLIPIcs

\graphicspath{{./figures/}}
\title{Reconfiguration of Externally Controlled Particles in Simple, Constructible Domains}

\author{Sándor P. Fekete}{Department of Computer Science, TU Braunschweig, Germany \and L3S Research Center, Hannover, Germany}{s.fekete@tu-bs.de}{https://orcid.org/0000-0002-9062-4241}{}
\author{Jonas Friemel}{Department of Electrical Engineering and Computer Science,\\ Bochum University of Applied Sciences, Germany}{jonas.friemel@hs-bochum.de}{https://orcid.org/0009-0009-6270-4779}{}
\author{Jan-Marc Reinhardt}{Department of Electrical Engineering and Computer Science,\\ Bochum University of Applied Sciences, Germany}{jan-marc.reinhardt@hs-bochum.de}{https://orcid.org/0009-0005-8907-3832}{}
\author{Christian Scheffer}{Department of Electrical Engineering and Computer Science,\\ Bochum University of Applied Sciences, Germany}{christian.scheffer@hs-bochum.de}{https://orcid.org/0000-0002-3471-2706}{}

\authorrunning{S. P. Fekete, J. Friemel, J.-M. Reinhardt, and C. Scheffer}

\Copyright{Sándor P. Fekete, Jonas Friemel, Jan-Marc Reinhardt, and Christian Scheffer}



\supplement{Videos illustrating our constructions can be found as follows:}
\supplementdetails[subcategory={Videos}]{Audiovisual}{https://doi.org/10.5281/zenodo.24558771}

\funding{Supported by the German Research Foundation (Deutsche Forschungsgemeinschaft, DFG), project SpaceAnts, 530918134.}

\nolinenumbers

\input{./include/custom-preamble}

\begin{document}

\maketitle

\begin{abstract}
  \input{./include/00-abstract}
\end{abstract}

\input{./include/01-introduction}
\input{./include/02-preliminaries}
\input{./include/03-permutations}

\input{./include/04-complexity}

\input{./include/05-conclusions}

\paragraph*{AI Statement}

No AI was used for any aspect of this paper.

\bibliography{references}

\clearpage

\appendix

\input{./include/0A-appendix}

\end{document}

%% file: include/custom-preamble.tex
\usepackage[basic]{complexity}
\usepackage{mathtools}
\usepackage{multirow}

\newcommand{\capref}[1]{\textsf{\textbf{\subref{#1}}}}

\newcommand*{\newterm}[1]{\emph{#1}}
\newcommand*{\probname}[1]{{\normalfont{\textsc{#1}}}}

\newcommand*{\bigO}{\mathcal{O}}
\newcommand*{\bigOtilde}{\tilde{\mathcal{O}}}
\newcommand*{\B}{\mathbb{B}}
\newcommand*{\Z}{\mathbb{Z}}
\newcommand*{\ZZ}{\Z^2}
\newcommand*{\N}{\mathbb{N}}
\newcommand*{\D}{\mathbb{D}}
\newcommand*{\moveseq}[1]{{\normalfont\textsc{#1}}}
\newcommand*{\udir}{{\normalfont\textsc{u}}}
\newcommand*{\ddir}{{\normalfont\textsc{d}}}
\newcommand*{\ldir}{{\normalfont\textsc{l}}}
\newcommand*{\rdir}{{\normalfont\textsc{r}}}
\newcommand*{\FT}{\mathrm{FT}}
\newcommand*{\SSt}{\mathrm{S1}}
\newcommand*{\IC}{\mathrm{IC}}
\newcommand*{\RR}{\mathrm{RR}}
\newcommand*{\NM}{\mathrm{NM}}
\newcommand*{\reconfig}{\probname{Reconfiguration}}
\newcommand*{\occupancy}{\probname{Occupancy}}
\newcommand*{\coccupancy}{\probname{Colored Occupancy}}
\newcommand*{\vacancy}{\probname{Vacancy}}

%% file: include/00-abstract.tex
In tilt models, large swarms of particles in grid environments move uniformly under the control of external signals such as magnetic fields.
We study particle reconfiguration problems in simple (hole-free) environments.
On the positive side, we design space-efficient simple environments to implement arbitrary permutations between labeled particle configurations, including a construction achieving worst-case optimal lengths of control sequences.
The environments we present are constructible via tilt assembly, where structures are built with bonding particles adhering to the same uniform movement rules as non-bonding particles.
To facilitate environment assembly, simplicity is a necessary condition.
Complementarily, we prove that established tilt problems like Occupancy and Vacancy are \PSPACE-complete in simple environments, settling an open question posed in~[BGC+, SODA~2020] and proving a conjecture of~[FKR+, ICALP~2025].

%% file: include/01-introduction.tex
\section{Introduction}
\label{sec:introduction}

The inherent restrictions imposed by micro- and nano-scale environments have led roboticists to investigate new control mechanisms.
Instead of individually steered robots, large swarms of robotic particles are manipulated by uniform external forces.
To deal with underactuation, it becomes crucial to utilize the geometric properties of the environment~\cite{tilt-video-socg15,tilt-iros13}.

\emph{Tilt models}, so called in reference to the image of objects sliding along a tilted surface under the effect of gravity, provide an abstract description amenable to algorithmic analysis.
They consider particles moving along axis-parallel paths in an environment specified by a bounded square grid.
In the \emph{full tilt model} ($\FT$), a control signal causes all particles to move maximally in a chosen direction until they are blocked by the boundary or another blocked particle, whereas the \emph{single step model} ($\SSt$) allows for precise unit steps.

One strain of research on tilt models considers fundamental problems posed for a \emph{given} environment, such as whether any particle can be moved to a specified position (\occupancy{}), or if there is a way to transform one of two given configurations of particles into the other (\reconfig).
Originally shown \NP-hard~\cite{tilt-nc19}, both problems are now known to be \PSPACE-complete in $\FT$~\cite{tilt-soda20}.
Another strain is concerned with the \emph{design} of environments to accomplish specific tasks, e.g., arranging particles in arbitrary shapes~\cite{tilt-reconfiguration-nc21}, sorting and classifying polyominoes~\cite{tilt-sort-classify}, or simulating bounded Turing machines~\cite{tilt-deterministic}.

Our interest lies in the influence that restrictions on the geometry of the environment have on these strains of research.
In particular, we want to know what changes when we only allow \emph{simple} environments, i.e., those that contain no holes.
Many environments engineered for particular applications rely on judiciously placed holes that serve as particle-blocking \emph{obstacles}.
Can the same tasks be accomplished in simple environments, maybe incurring an overhead of some kind?
The computational complexity of \occupancy{} and \reconfig{} in simple environments was posed as an open problem in~\cite{tilt-soda20}.

On the one hand, the distinction between simple and non-simple environments is interesting for purely geometric reasons.
Proving a problem hard can be easier for environments with holes.
It took more than 40 years from the proof that packing $(2 \times 2)$ squares into rectilinear polygons is \NP-hard~\cite{nonsimple-packing-hard}, until the result was extended to simple polygons~\cite{simple-packing-hard}.
In contrast, some problems become manageable when the geometry is restricted.
The famous Watchman problem is hard in general~\cite{nonsimple-watchman-hard}, but becomes tractable in simple polygons~\cite{simple-watchman-easy}.

On the other hand, simple environments offer the distinct advantage of viable constructibility.
Existing work on environment design typically does not take into account how to construct the environment.
Yet, the solution may be found within the same body of work in the form of \emph{tilt assembly}~\cite{assembly-algorithmica20}, which examines how to build structures with particles that bond on contact.
We propose to build environments via bonding particles before using them for tasks involving other kinds of (non-bonding) particles.
Although not every simple environment is constructible by established methods, simplicity is a necessary condition.

\begin{table}
  \caption{%
    Results for $c$-colored or labeled $(w \times h)$-matrix permutations with $k = wh$ particles.
  }\label{tbl:summary}%
  \centering%
  \begin{tabular}{lccllll}%
    \hline
    & Env.   & Colors & Steps              & Bounding Box                        & Obstacles         & Reference \\
    \hline\hline
    \multirow{4}{*}{\rotatebox{90}{Fixed\ }}
    & Holes  & $k$    & $4$                & $\bigO(k^2)$                        & $\bigO(k)$        & \cite[Thm.~2]{tilt-nc19} \\
    & Holes  & $k$    & $11$               & $\bigO(k \cdot (w + h))$            & $\bigO(k)$        & \cite[Sec.~IV]{tilt-rearranging} \\
    & Simple & $k$    & $\bigO(\log k)$    & $\bigO(k \log k + k \min \{w, h\})$ & $\bigO(k \log k)$ & \textbf{Cor.~\ref{cor:fixed-k-perm}, Thm.~\ref{thm:permutation-assembly}} \\
    & Simple & $c$    & $\bigO(\log c)$    & $\bigO(k^2 \log c)$                 & $\bigO(k \log c)$ & \textbf{Thms.~\ref{thm:fixed-c-perm}, \ref{thm:permutation-assembly}} \\
    \hline
    \multirow{4}{*}{\rotatebox{90}{Arbitrary\ }}
    & Holes  & $k$    & $\bigO(k \log k)$  & $\bigO(k^3)$                        & $\bigO(k^2)$      & \cite[Thm.~8]{tilt-nc19} \\
    & Holes  & $k$    & $\bigO(k^2)$       & $\bigO(k)$                          & $\bigO(w + h)$    & \cite[Sec.~V]{tilt-rearranging} \\
    & Simple & $c$    & $\bigO(k \log c)$  & $\bigO(k \min \{w, h\})$            & $\bigO(k)$        & \textbf{Thms.~\ref{thm:arbitrary-c-perm-fast}, \ref{thm:permutation-assembly}} \\
    & Simple & $c$    & $\bigO(k^2)$       & $\bigO(k)$                          & $\bigO(w + h)$    & \textbf{Thms.~\ref{thm:arbitrary-c-perm-bubble}, \ref{thm:permutation-assembly}} \\
    \hline
  \end{tabular}%
\end{table}

\subsection*{Our contributions}

We provide new results on the reconfiguration of particles in simple environments using external control signals, principally within the full tilt model of uniform control.

On the algorithmic side, we design simple, constructible environments for \emph{matrix permutations}, i.e., rearranging rectangular configurations of colored particles into new configurations of the same shape but with permuted particle positions.
We distinguish between the case of a known, fixed permutation and constructions admitting all permutations via distinct control sequences.
See Table~\ref{tbl:summary} for a summary of the performance in terms of space requirements and lengths of control sequences compared to existing work using non-simple environments.
Notably, while we do not match the constant-length sequences achieved for fixed permutations in non-simple environments, our constructions for arbitrary permutations match or improve upon existing results and achieve worst-case optimal sequence lengths (Observation~\ref{obs:lower-bound}).

As a second major contribution, we settle the complexity of \reconfig{} and \occupancy{} in simple environments by providing a novel reduction from \probname{Finite Function Generation}~\cite{kozen}, showing them \PSPACE-complete (Corollary~\ref{cor:hardness-occupancy-ft}).
An~extension of the reasoning allows us to prove the conjecture~\cite{tilt-fill-icalp} that \vacancy{}, the problem to determine if a given position can be freed, is \PSPACE-complete even in the single step model (Theorem~\ref{thm:vacancy-hard-sst}).

A~collection of videos illustrating our constructions can be found at~\url{https://zenodo.org/records/24558771}.
Full proofs of most statements have been deferred to Appendix~\ref{sec:appendix}.

\subsection*{Related work}

Tilt models were introduced and motivated from a practical perspective in~\cite{tilt-iros13}.
Many foundational theoretical results for~$\FT$ have been collected in~\cite{tilt-nc19}, including \NP-hardness of fundamental problems and first results on matrix permutations.
The latter, along with space-efficiency improvements from~\cite{tilt-rearranging}, are depicted in comparison to our results in Table~\ref{tbl:summary}.

Further investigation~\cite{tilt-soda20,tilt-soda19,tilt-automata-socg} into the complexity of \reconfig{} and \occupancy{} in~$\FT$ has revealed that both are \PSPACE-complete in general and \NP-hard for simple environments.
In contrast, \occupancy{} is easy in~$\SSt$~\cite{tilt-limited-directions-jip20}, but \probname{Relocation}, the question if a particular particle can be moved to a chosen position, remains \PSPACE-complete~\cite{tilt-hardness-cccg20}.
If one limits the number of directions, \probname{Relocation} in~$\SSt$ is \NP-complete, even within squares~\cite{tilt-limited-directions-nc25}.

A~major idea is to assemble bonding particles into larger structures using external control.
Initially proposed in~\cite{assembly-ral17}, early algorithmic work on assembly in $\FT$~\cite{assembly-algorithmica20} was followed by parallel assembly~\cite{assembly-parallel}, assembly via single-step moves~\cite{assembly-single-step}, universal constructors that can build arbitrary shapes from certain classes~\cite{tilt-soda19}, the drop shape hierarchy of constructible shapes~\cite{tilt-soda20}, and experimental evaluation of heuristics~\cite{assembly-iros23}; see also the discussion in Section~\ref{sec:permutations-assembly}.

Other goals include the gathering of particles~\cite{gathering-icra20,tilt-automata-socg,gathering-rl-iros22,gathering-case16}, mapping an unknown environment~\cite{tilt-mapping}, and filling/draining~\cite{tilt-fill-icalp}.
Environments have been designed to sort and classify polyominoes~\cite{tilt-sort-classify}, reconfigure particles into arbitrary shapes~\cite{tilt-reconfiguration-nc21}, and simulate computations by deterministic control sequences~\cite{tilt-deterministic}.

Various models of motion planning share similarities with tilt models.
The rules of the game ``2048''~\cite{2048-fun16,2048-cccg20} resemble the full tilt model in a square, where the single step model corresponds to blocks being pushed by sweeping lines~\cite{sweeping-eurocg16,sweeping-cgt23}.
Apart from the discrete grids of tilt models, uniform control in continuous settings has been investigated in the context of robot localization~\cite{localization-tor07} and draining liquids from polygons~\cite{draining-a-polygon}.
The maximal movement of~$\FT$ augmented with individual control appears in several games~\cite{randolph-endm06,randolph-jip17,atomix}.

There is a long history of \PSPACE-completeness results for geometric reconfiguration, with variously shaped moving pieces, e.g.,~\cite{simple-square-reconfiguration-pspace,rush-hour-unit-fixed-pspace,rush-hour-pspace,sliding-block-pspace,warehouseman-pspace}.
Our result in Corollary~\ref{cor:hardness-occupancy-ft} is noteworthy for having both restricted geometry and only very simple shapes (unit squares).

The reconfiguration of colored unit-size particles on a square grid has also been considered in the context of \emph{coordinated} motion planning, where particles move in parallel but possibly in different directions~\cite{reconfig-bounded-stretch,reconfig-densely-packed}.
There, it is generally easier to decide if reconfiguration is possible, and one can instead look for (approximately) shortest control plans.

%% file: include/02-preliminaries.tex
\section{Preliminaries}
\label{sec:preliminaries}

\begin{figure}
  \centering%
  \begin{subcaptionblock}{53pt}%
    \centering%
    \includegraphics{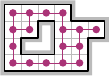}%
    \caption{}%
    \label{fig:environment-example}%
  \end{subcaptionblock}%
  \hfil%
  \begin{subcaptionblock}{237pt}%
    \centering%
    \includegraphics{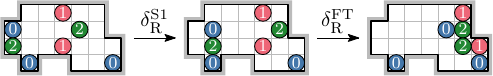}%
    \caption{}%
    \label{fig:move-example}%
  \end{subcaptionblock}%
  \caption{%
    \capref{fig:environment-example}~A non-simple environment and its boundary with highlighted nodes.
    \capref{fig:move-example}~Move~$\rdir$ in the single step ($\SSt$) and full tilt ($\FT$) models in a simple environment with a $3$-colored configuration.
  }%
\end{figure}

Consider the infinite unit grid graph~$G$ with integer coordinates, where two nodes $u, v \in \ZZ$ share an edge exactly if $\|u - v\|_1 = 1$.
An \newterm{environment}~$B = (V, E)$ is defined as a finite connected subgraph of the integer grid~$G$ induced by $V \subsetneq \ZZ$, see Figure~\ref{fig:environment-example}.
The \newterm{boundary} of an environment~$B$ is the unique axis-aligned polygon with integer side lengths separating~$V$ from $\ZZ \setminus V$.
Each node~$v \in V$ is the intersection of two \newterm{segments}, a \newterm{row segment} and a \newterm{column segment}, which are maximally contiguous subgraphs of~$B$ with equal $y$- and $x$-coordinates, respectively.
An environment is \newterm{simple} if it is hole-free, i.e., if $\ZZ \setminus V$ is connected.

For a given environment, a \newterm{configuration} is defined by a map $C : V \to \N \cup \{\bot\}$, where a node $v \in V$ is \newterm{free} if $C(v) = \bot$ and otherwise \newterm{occupied} by a particle with \newterm{label} $C(v) \in \N$.
We denote the number of particles in a configuration~$C$ by $k \coloneq |C^{-1}(\N)|$ and the number of labels in~$C$ by $c \coloneq |C(V) \setminus \{\bot\}|$.
A configuration~$C$ with $c$ labels is \newterm{$c$-colored}.
In adherence to established terminology, we refer to a $1$-colored configuration as \newterm{unlabeled} and to a $k$-colored configuration as \newterm{labeled}.
See Figure~\ref{fig:move-example} for a $3$-colored configuration.
For any label~$\ell \in \N$, the \newterm{multiplicity of $\ell$} is the number of particles with this label, i.e., $|C^{-1}(\ell)|$.
In an unlabeled configuration, there is a unique label with multiplicity~$k$, while in a labeled configuration, all labels have multiplicity~$0$ or~$1$.
Two configurations $C_1, C_2$ with the same number of particles are \newterm{compatible} if they have matching multiplicites for all labels.

For $\ell \in \N$, we use $[\ell]$ to denote the set of integers $\{0, \dots, \ell - 1\}$.
We assume without loss of generality that all labels~$\ell$ in a $c$-colored configuration~$C$ satisfy $0 \leq \ell < c$, i.e., $C(V) \setminus \{\bot\} = [c]$; we can do so, because it is always possible to consistently rename labels.

We denote the four movement directions (up, down, left, and right) by $\udir$, $\ddir$, $\ldir$, and~$\rdir$, and write $\D \coloneq \{\udir, \ddir, \ldir, \rdir\}$.
For each direction~$d \in \D$, a tilt model~$M$ defines a \newterm{move} $\delta_d^M : (V \to \N \cup \{\bot\}) \to (V \to \N \cup \{\bot\})$ that maps between compatible configurations.
We inductively extend moves to arbitrary (possibly empty) sequences of directions by setting $\delta_{wd}^M = \delta_d^M \circ \delta_w^M$, $d \in \D$ and $w \in \D^\ast$, and $\delta_\varepsilon^M = \operatorname{id}$, where~$\varepsilon$ is the empty sequence and~$\operatorname{id}$ the identity function.
If the model~$M$ is clear from context, we identify a move~$\delta_w^M$~by~its~direction~sequence~$w \in \D^\ast$.

We first define the move~$\delta_\ldir^{\SSt}$ in the single step model~$\SSt$.
Other directions are analogous.
Let $v \in V$ be a node, let $\ell \in \N$ be a label, and let $C_1, C_2$ be (compatible) configurations such that $\delta_\ldir^{\SSt}(C_1) = C_2$.
Then $C_2(v) = \ell$ if either (i)~$C_1(v) = \ell$ and all nodes to the left of $v$ in its row segment are occupied, or (ii)~$C_1(u) = \ell$ for the right neighbor~$u$ of~$v$ and one of the nodes to the left of~$u$ in its row segment is free.
If $C_2(v) \neq \ell$ for all $\ell \in \N$, then $C_2(v) = \bot$.

In the full tilt model~$\FT$, particles move maximally until they are blocked.
Following this intuition, we extend the definition of move~$\delta_\ldir^{\SSt}$ to $\delta_\ldir^{\FT}$.
Let $C_1$ be a configuration.
We recursively define $C_i \coloneq \delta_\ldir^{\SSt}(C_{i - 1})$.
There is a finite~$j \in \N$ such that $C_j = C_{j - 1}$.
Then $\delta_\ldir^{\FT}(C_1) \coloneq C_j$ for any initial configuration~$C_1$ and $j \in \N$ such that $C_j = C_{j - 1}$.
In other words, $\delta_\ldir^{\FT}$ is the fixed point of composition with~$\delta_\ldir^{\SSt}$ reached by iterating~$\delta_\ldir^{\SSt}$.
See Figure~\ref{fig:move-example}.

%% file: include/03-permutations.tex
\section{Permutations of rectangular particle arrangements}
\label{sec:permutations}

We refer to a map~$M: H \to \N$ for some finite rectangular subgraph of the integer grid~$G$ with dimensions $w \times h$ and node set~$H \subsetneq \ZZ$ as a \newterm{$(w \times h)$-matrix}.
If~$w$ and~$h$ are clear from the context, $M$ is simply a \newterm{matrix}.
If $H$ is a subgraph of some environment $B = (V, E)$, then the extension of~$M$ to~$V$ with $v \mapsto \bot$ for $v \in V \setminus H$ defines a configuration of~$B$.
Other definitions for configurations apply to matrices analogously.
In particular, we refer to the members of $M(H)$ as \newterm{labels} and two matrices~$M_1$ and~$M_2$ with the same domain are \newterm{compatible} if the multiplicities of their labels match.
In a given environment, a move sequence that transforms a matrix~$M_1$ into a compatible matrix~$M_2$ is called a \newterm{matrix permutation}.
For $c < wh = k$, these permutations are multiset permutations as some labels~occur~multiple~times.

\input{./include/03-1-fixed}
\input{./include/03-2-arbitrary}
\input{./include/03-3-assembly}

%% file: include/03-1-fixed.tex
\subsection{Environments for fixed permutations}
\label{sec:permutations-fixed}

For any pair of compatible matrices~$M_s$ and~$M_t$, it is possible to create an environment such that there is a matrix permutation of constant length that transforms~$M_s$ into~$M_t$~\cite{tilt-nc19,tilt-rearranging}.
However, all existing constructions make use of precisely positioned, disconnected obstacles.
Thus, the resulting environment are inherently non-simple.
We construct simple environments that enable matrix permutations of length $\bigO(\log c)$ for given compatible $c$-colored matrices.

We first consider labeled, i.e., $k$-colored, $(k \times 1)$-matrices before extending our construction to permutations of $c$-colored $(w \times h)$-matrices.
We introduce a linear-size gadget that partitions particles into two groups in four moves without changing the relative order of particles within a partition.
By chaining $\lceil\log k\rceil$ versions of this gadget together, we can realize any fixed matrix permutation via radix sort in $\bigO(\log k)$ moves.

\begin{restatable}{theorem}{restateFixedFlatPerm}
  \label{thm:fixed-flat-k-perm}
  Let $M_s$ and $M_t$ be compatible labeled $(k \times 1)$-matrices.
  Then there exists a simple environment~$B$ in a bounding box of size~$\bigO(k \log k)$ such that~$M_s$ can be transformed into~$M_t$ by a matrix permutation of length~$\bigO(\log k)$ in~$B$.
\end{restatable}
\begin{proof}[Proof sketch]
  Consider a (non-matrix) permutation~$\pi$ of $[k]$.
  To sort~$\pi$ via radix sort, elements are recursively partitioned bitwise, starting with the least-significant bit.
  In each round, relative order is maintained within the partitions.
  After $\lceil \log k \rceil$ partitioning rounds, the permutation is sorted~\cite{taocp-vol3}.
  This approach can be used to reconfigure between arbitrary pairs of permutations of the same integers via relabeling.
  To emulate the partitioning rounds, we construct one subset partitioning gadget per bit of~$k$, see Figure~\ref{fig:fixed-labeled-subset-picker}.
  \begin{figure}[tb]
    \centering%
    \begin{subcaptionblock}{209pt}%
      \centering%
      \includegraphics[page=1]{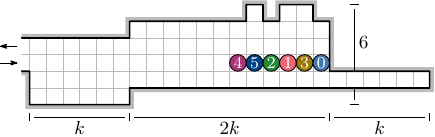}%
      \caption{}%
      \label{fig:fixed-labeled-subset-picker-a}%
      \vspace{0.3cm}%

      \includegraphics[page=2]{permutations/fixed-labeled-subset-picker}%
      \caption{}%
      \label{fig:fixed-labeled-subset-picker-b}%
    \end{subcaptionblock}%
    \hfill%
    \begin{subcaptionblock}{186pt}%
      \centering%
      \includegraphics{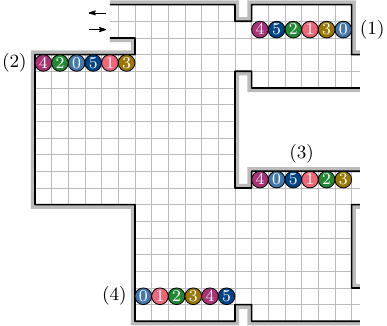}%
      \caption{}%
      \label{fig:fixed-labeled-permutation}%
    \end{subcaptionblock}
    \caption{%
      \capref{fig:fixed-labeled-subset-picker-a}~A subset partitioning gadget as used in the construction of Theorem~\ref{thm:fixed-flat-k-perm} for $k = 6$ that moves particles with even labels to the front.
      Particles ordered $4\,5\,2\,1\,3\,0$ enter the gadget with~$(\ddir) \rdir$.
      \capref{fig:fixed-labeled-subset-picker-b}~After~$\moveseq{ul}$, particles with even labels are separated from particles with odd labels.
      With~$(\moveseq{drul})^2$, the particles leave the gadget ordered $4\,2\,0\,5\,1\,3$.
      \capref{fig:fixed-labeled-permutation}~Chaining $\lceil \log_2 6 \rceil = 3$ subset partitioning gadgets on the right for sets $\{0,2,4\}$, $\{0,1,4,5\}$, and $\{0,1,2,3\}$ allows us to perform a labeled matrix permutation for six particles in $3 \cdot 3 + 2 = 11$ repetitions of~$\moveseq{drul}$.
      The figure~shows~intermediate~configurations.
    }%
    \label{fig:fixed-labeled-subset-picker}%
  \end{figure}
\end{proof}

\begin{figure}[tb]
    \centering%
    \includegraphics{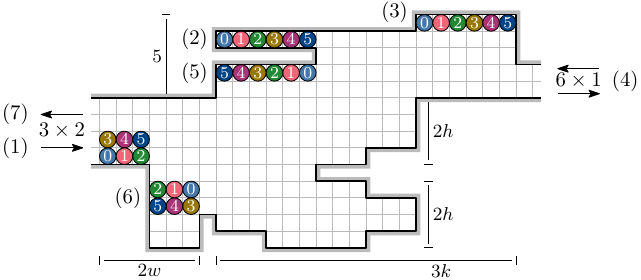}%
    \caption{%
      The flattening gadget for converting between $(3 \times 2)$- and $(6 \times 1)$-matrices.
      A sorted $(3 \times 2)$-matrix enters the gadget (1), is flattened into a $(6 \times 1)$-matrix (2), and is transported into the $(6 \times 1)$-permutation construction from Theorem~\ref{thm:fixed-flat-k-perm} (3--4).
      The matrix returns reversed (5), is turned back into a $(3 \times 2)$-matrix (6) and is transported out of the gadget (7).
    }%
    \label{fig:matrix-flattener}%
  \end{figure}

The construction above can be prepended with a gadget to turn a $(w \times h)$-matrix into a $(k \times 1)$-matrix and vice versa, see Figure~\ref{fig:matrix-flattener}, yielding the following result.

\begin{restatable}{corollary}{restateFixedKPerm}
  \label{cor:fixed-k-perm}
  Let $M_s$ and $M_t$ be compatible labeled $(w \times h)$-matrices.
  Then there exists a simple environment~$B$ in a bounding box of size~$\bigO(k \log k + k \min \{w, h\})$ such that~$M_s$ can be transformed into~$M_t$ by a matrix permutation of length~$\bigO(\log k)$ in~$B$.
\end{restatable}

This approach extends to $c$-colored instances for any $c \leq k$ in $\bigO(\log k)$ steps.
But faster permutations in $\bigO(\log c)$ steps are more tricky as it is no longer possible to relabel particles for reconfiguration via sorting.
Instead, we first sort the particles without relabeling in $\bigO(\log c)$ steps and then un-sort to the target matrix using new inverse subset partitioning~gadgets.

\begin{restatable}{theorem}{restateFixedCPerm}
  \label{thm:fixed-c-perm}
  Let $M_s$ and $M_t$ be compatible $c$-colored $(w \times h)$-matrices.
  Then there exists a simple environment~$B$ in a bounding box of size~$\bigO(k^2 \log c)$ such that~$M_s$ can be transformed into~$M_t$ by a matrix permutation of length~$\bigO(\log c)$ in~$B$.
\end{restatable}
\begin{proof}[Proof sketch]
  To invert a subset partitioning step, we construct a gadget that can individually interleave the $q$ rightmost particles of a $(k \times 1)$-matrix between the $k - q$ leftmost particles without changing their relative order.
  The gadget consists of two stages.
  First, the $k - q$ leftmost particles are reversed, see Figure~\ref{fig:inverse-subset-picker-invert}.
  Then, the~$q$ rightmost particles are interleaved between the $k - q$ leftmost particles, which also restores the original relative order of these particles, see Figure~\ref{fig:inverse-subset-picker-insert}.
  Now, for both~$M_s$ and~$M_t$, we compute sorting sequences via radix sort.
  We first chain subset partitioning gadgets to sort~$M_s$.
  Then, we invert the sequence computed for sorting~$M_t$ and chain the corresponding inverse subset partitioning gadgets.%
  \begin{figure}
    \begin{subcaptionblock}{185pt}%
      \centering%
      \includegraphics{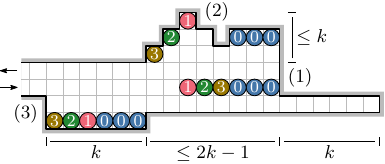}%
      \caption{}%
      \label{fig:inverse-subset-picker-invert}%
    \end{subcaptionblock}%
    \hfill%
    \begin{subcaptionblock}{179pt}%
      \centering%
      \includegraphics{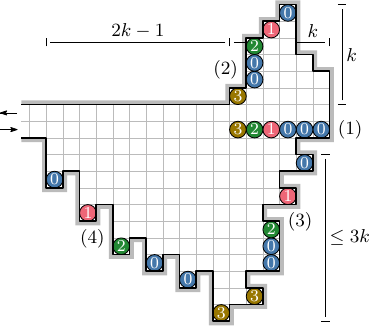}%
      \caption{}%
      \label{fig:inverse-subset-picker-insert}%
    \end{subcaptionblock}%
    \caption{%
      An inverse subset partitioning gadget from Theorem~\ref{thm:fixed-c-perm} that works in two stages.
      \capref{fig:inverse-subset-picker-invert}~The first stage.
      Particles ordered $1\,2\,3\,0\,0\,0$ enter the gadget (1), the first three particles are separated from the rest and inverted with moves $\moveseq{ul}$ (2), and the particles are re-concatenated with moves $\moveseq{druld}$ (3).
      \capref{fig:inverse-subset-picker-insert}~The second stage.
      The particles from the first stage enter the gadget (1), the three $0$-particles on the right are interleaved between the other three particles with moves $\moveseq{ul}$ (2), and the particles are re-flattened with moves $\moveseq{dr}$ (3) and $\moveseq{uld}$ (4) to exit the gadget ordered $0\,1\,2\,0\,0\,3$~with~$\moveseq{rul}$.
    }%
    \label{fig:inverse-subset-picker}%
  \end{figure}
\end{proof}

On first glance, some of the gadgets may appear needlessly complicated.
For example, the rightmost $k$-node nooks of Figures~\ref{fig:fixed-labeled-subset-picker-a} and~\ref{fig:inverse-subset-picker-invert} serve no apparent purpose.
The reason behind these design choices is pipelining.
For each of the gadgets, the sequence to move a particle configuration from the gadget's entry to its exit is $(\moveseq{drul})^\ell$ for some constant~$\ell \in \N$.
This allows us to apply multiple permutations in parallel with a constant move offset between each particle set without interference.
In particular, applying the same permutation to~$j$~different particle groups in a row is possible in $\bigO(j + \log c)$ steps~instead~of~$\bigO(j \log c)$.

%% file: include/03-2-arbitrary.tex
\subsection{Arbitrary permutations in a single environment}
\label{sec:permutations-arbitrary}

All previous environments were constructed to enable efficient permutations for a given pair $(M_s, M_t)$ of compatible matrices.
We now focus our attention on general-purpose environments in which a given matrix~$M$ can be transformed into any compatible matrix.

\begin{restatable}{theorem}{restateArbitraryFast}
  \label{thm:arbitrary-c-perm-fast}
  Let~$M$ be a $c$-colored $(w \times h)$-matrix.
  Then there exists a simple environment~$B$ in a bounding box of size $\bigO(k \min \{w, h\})$ such that~$M$ can be transformed into any compatible matrix by a matrix permutation of length~$\bigO(k \log c)$ in~$B$.
\end{restatable}
\begin{proof}[Proof sketch]
  We design an environment for splitting the particles into two stacks, retaining their relative order within the stacks.
  The particles can be repeatedly popped from either of the two stacks, which allows us to emulate the radix-based strategy for fixed permutations from Theorem~\ref{thm:fixed-c-perm} in $\bigO(k \log c)$ steps for any compatible permutation.
  See Figure~\ref{fig:two-stacks}.
  \begin{figure}
    \centering%
    \includegraphics{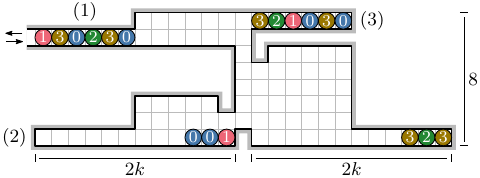}%
    \caption{%
      The two-stack construction for arbitrary matrix permutations from Theorem~\ref{thm:arbitrary-c-perm-fast} with $k = 6$ particles.
      First, the particles enter the gadget ordered $1\,3\,0\,2\,3\,0$ (1).
      They are separated into two stacks (left $0$ and $1$, right $2$ and $3$) via moves $\moveseq{rd ldrd rd rd ldrd rd ldrd}$ (2).
      Then, the particles are popped from the two stacks with moves $(\moveseq{lulurd})^2\ (\moveseq{rurdlulurd})^2\ \moveseq{lulurd rurdlulurd}$ (stacks right, right, left, left, right, left), resulting in order $3\,2\,1\,0\,3\,0$ (3).
    }%
    \label{fig:two-stacks}%
  \end{figure}
\end{proof}

This construction matches the worst-case permutation length achieved for labeled matrices in~\cite{tilt-nc19}, but without the need for holes and with significantly lower space requirements.
In fact, a worst-case permutation length of $\bigO(k \log c)$ is optimal.

\begin{restatable}{observation}{restateLowerBound}\label{obs:lower-bound}
  Let~$M$ be a $c$-colored matrix and let~$B$ be an environment where~$M$ can be transformed into any compatible matrix by a matrix permutation.
  Then there is a compatible matrix $M'$ such that any permutation transforming~$M$ into~$M'$ in~$B$ has length~$\Omega(k \log c)$.
\end{restatable}

For scenarios where neither~$w$ nor~$h$ are constant and environment compactness has a higher priority than low permutation length, we additionally present a linear-space construction that enables arbitrary matrix permutations in a quadratic number of moves via a bubble-sort-type strategy.
The linear-space construction by Zhang, Cheng, Qi, and Balkcom~\cite{tilt-rearranging} uses the same idea, albeit with holes.
To achieve the space savings, we manipulate $(w \times h)$-matrices directly instead of using intermediate $(k \times 1)$ transformations.

\begin{restatable}{theorem}{restateBubbleSort}
  \label{thm:arbitrary-c-perm-bubble}
  Let~$M$ be a $(w \times h)$-matrix.
  Then there exists a simple environment~$B$ in a bounding box of size $\bigO(k)$ such that~$M$ can be transformed into any compatible matrix by a matrix permutation of length~$\bigO(k^2)$ in~$B$.
\end{restatable}

%% file: include/03-3-assembly.tex
\subsection{Assembling permutation constructions}
\label{sec:permutations-assembly}

Up to this point, we have only examined the \newterm{blocking} variant of the full tilt model, where particles block each other when colliding, but do not interact further.
A separate line of research, originating with~\cite{assembly-ral17}, considers the \newterm{assembly} variant.
Here, particles form persistent bonds as soon as they become adjacent, which enables constructing multi-particle shapes.
We refer to such bonding particles as \newterm{assembly particles}.

When examining the constructibility of shapes, it is typically assumed that a configuration is initialized with an immovable \newterm{seed} assembly particle and subsequent assembly particles are sequentially ``dropped'' onto the constructed shape from one of the four compass directions~\cite{assembly-algorithmica20}.
Shapes that can be constructed in this way are referred to as \newterm{drop shapes}~\cite{tilt-soda19}.
Such shapes can, in turn, be combined in the same manner to create more complex shapes~\cite{assembly-parallel}.
This idea naturally leads to a \newterm{drop shape hierarchy~$\mathcal{D}$}~\cite{tilt-soda20}.
The first level consists of all drop shapes made up of individual assembly particles as defined above.
All subsequent levels are defined recursively:
Level~$i$ consists of all drop shapes that can be constructed by sequentially dropping shapes from level~$i - 1$ onto a seed configuration from level~$i - 1$.

\begin{figure}
  \begin{subcaptionblock}{61pt}%
    \centering%
    \includegraphics[page=1]{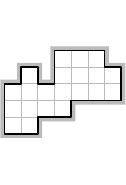}%
    \caption{}%
    \label{fig:assembly-example-env}%
  \end{subcaptionblock}%
  \hfill%
  \begin{subcaptionblock}{89pt}%
    \centering%
    \includegraphics[page=2]{assembly/example}%
    \caption{}%
    \label{fig:assembly-example-order}%
  \end{subcaptionblock}%
  \hfill%
  \begin{subcaptionblock}{61pt}%
    \centering%
    \includegraphics[page=1]{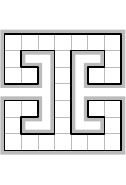}%
    \caption{}%
    \label{fig:assembly-not-constructible-env}%
  \end{subcaptionblock}%
  \hfill%
  \begin{subcaptionblock}{89pt}%
    \centering%
    \includegraphics[page=2]{assembly/not-constructible}%
    \caption{}%
    \label{fig:assembly-not-constructible-particles}%
  \end{subcaptionblock}%
  \caption{%
    \capref{fig:assembly-example-env}~A constructible simple environment~$B$.
    \capref{fig:assembly-example-order}~The environment's assembly shape~$A(B)$ and a valid assembly order to construct $A(B)$ as a drop shape with seed particle~$0$.
    Red assembly particles are dropped from the right (1--11) and blue particles from the left (12--28).
    \capref{fig:assembly-not-constructible-env}~A simple environment that is not constructible on the first level of $\mathcal{D}$.
    For each possible seed, at least one of the four cyan positions in \capref{fig:assembly-not-constructible-particles} becomes before an assembly particle can be dropped there.
  }%
\end{figure}

We identify a drop shape~$D$ by the coordinates $D \subsetneq \ZZ$ of nodes occupied by assembly particles.
Given an environment~$B = (V, E)$, we refer to the node set $A(B) \coloneq \{(x + i, y + j) : (x, y) \in V, i,j \in \{-1, 0, 1\}\} \setminus V$ as the \newterm{obstacle shape of $B$}, where each node $v \in A(B)$ is an \newterm{obstacle}.
The environment~$B$ is \newterm{constructible} (on level~$i$ of~$\mathcal{D}$) if $A(B)$ is a drop shape (on level~$i$ of~$\mathcal{D}$), see Figures~\ref{fig:assembly-example-env}\nobreakdash--\subref{fig:assembly-example-order}.
By this definition, a constructible environment is necessarily simple;
otherwise, the corresponding obstacle shape would need to be disconnected.
Therefore, permutation environments from prior work, which contain holes~\cite{tilt-nc19,tilt-rearranging}, are not constructible as the corresponding obstacle shapes contain disconnected obstacles.
However, simplicity is not sufficient to guarantee constructability, see Figures~\ref{fig:assembly-not-constructible-env}\nobreakdash--\subref{fig:assembly-not-constructible-particles}.

We show that the permutation environments presented earlier are constructible.

\begin{restatable}{theorem}{restatePermutationAssembly}
  \label{thm:permutation-assembly}
  The environments for fixed and arbitrary $c$-colored $(w \times h)$-matrix permutations from Theorems~\ref{thm:fixed-c-perm}, \ref{thm:arbitrary-c-perm-fast}, and~\ref{thm:arbitrary-c-perm-bubble} are constructible on the first level of~$\mathcal{D}$.
  The corresponding obstacle shapes consist of $\bigO(k \log c)$, $\bigO(k)$, and $\bigO(w + h)$ obstacles, respectively.
\end{restatable}
\begin{proof}[Proof sketch]
  We show that two-stack construction for arbitrary $(k \times 1)$-matrices from Theorem~\ref{thm:arbitrary-c-perm-fast}, see Figure~\ref{fig:two-stacks}, is constructible.
  The constructibility of the other environments is proven analogously.
  The assembly shape and assembly order for this construction with $k = 6$ is shown in Figure~\ref{fig:assemble-two-stacks}.
  For larger values of~$k$, it suffices to cut the assembly shape vertically along the four dashed lines and insert assembly particles in the resulting gaps.
  Counting the number of steps reveals that the obstacle shape consists of $13k + 22 = \bigO(k)$ obstacles.
  \begin{figure}
    \centering%
    \includegraphics{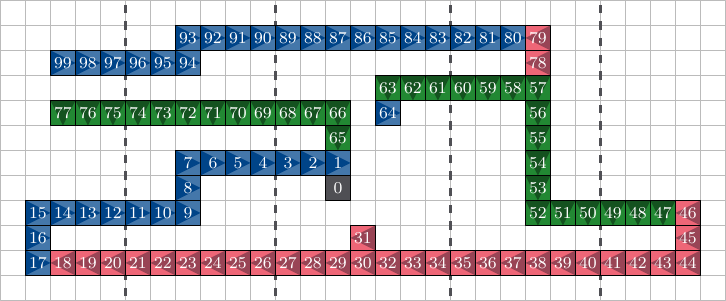}%
    \caption{%
      Assembly order for the two-stack construction from Figure~\ref{fig:two-stacks} with seed particle~$0$.
      Blue assembly particles are dropped from the left (1--17, 64, and 80--99), red particles from the right (18--46 and 78--79), and green particles from the top (47--63 and 65--77).
      The dashed lines show where to add or remove assembly particles to extend the construction to other matrix dimensions.
    }%
    \label{fig:assemble-two-stacks}%
  \end{figure}
\end{proof}

%% file: include/04-complexity.tex
\section{Computational complexity of tilt problems in simple domains}
\label{sec:complexity}

We turn our attention to the computational complexity of fundamental tilt problems.
Given an environment~$B$ and two configurations~$C_1, C_2$ of~$B$, \reconfig{} is the problem to determine if there is a sequence~of~moves~$w$ such that $\delta_w(C_1) = C_2$.
\coccupancy{} asks, given an environment~$B$, a configuration~$C$ of~$B$, a node~$p$, and a label~$\ell \in \N$, if there is a sequence~of~moves~$w$ such that $\delta_w(C)(p) = \ell$.
The restrictions of \coccupancy{} to labeled and unlabeled configurations are called \probname{Relocation} and \occupancy{}, respectively.
When the authors of~\cite{tilt-soda20} proved these problems \PSPACE-complete for general environments in $\FT$, they posed the complexity for simple environments as an open problem.
Later work~\cite{tilt-automata-socg} established \NP-hardness but did not determine the exact complexity class.

\input{./include/04-1-simple}
\input{./include/04-2-vacancy}

%% file: include/04-1-simple.tex
\subsection{PSPACE-completeness in simple domains}
\label{sec:complexity-simple}

The main goal of this subsection is to prove the following theorem, in which the \newterm{leftmost column} refers to the subset of nodes with minimum $x$-coordinate.
Afterwards, we adapt the proof to show \PSPACE-hardness of various tilt problems for simple environments.

\begin{restatable}{theorem}{restateHardnessMainThm}\label{thm:hardness-main-thm}
It is \PSPACE-hard to decide, given a simple environment~$B$ and a configuration~$C$ of $B$, if there is a sequence~of~moves~$w \in \D^*$ such that only nodes in the leftmost column of~$B$ are occupied in~$\delta^{\FT}_w(C)$.
\end{restatable}

For a collection~$F$ of transformations on the same domain~$S$, we recursively define $F^*$ to contain the identity on~$S$ and every composition $f \circ g$ for $f \in F$ and $g \in F^*$, i.e., $F^*$ contains all compositions of functions from~$F$.
A function~$h \in F^*$ is \newterm{generated} by~$F$.
We employ a reduction from \probname{Finite Function Generation (FFG)}: Given a number~$m \in \N$, a family $f_i: [m] \to [m]$ of $j > 0$ functions, $0 \leq i < j$, and a function $h: [m] \to [m]$, is $h$ generated by $\{f_i : i \in [j]\}$?
\probname{FFG} has been proven \PSPACE-complete by Kozen~\cite{kozen}.
As an intermediate step, we reduce \probname{FFG} to \probname{FFG-2}, its restriction to a family of $j=2$ functions:
Given a number~$m \in \N$ and three transformations $f, g, h: [m] \to [m]$, is $h$ generated by~$\{f,g\}$?

While we are unaware of previous use of this variant in the literature, the reduction from \probname{FFG} to \probname{FFG-2} is straightforward and builds on established techniques.

\begin{restatable}{lemma}{restateFFGhard}\label{lem:ffg2-hard}
\probname{FFG-2} is \PSPACE-complete.
\end{restatable}

We present a polynomial-time procedure turning a given instance of \probname{FFG-2} with functions $f,g,h$ into a simple environment~$B$ with a configuration whose particles can all be brought to the leftmost column of~$B$ if and only if $h$ is generated by~$\{f,g\}$.
For ease of exposition, we proceed in several steps and first allow holes before showing how to eliminate them.

\begin{figure}[tb]
  \centering%
  \begin{subcaptionblock}{110pt}%
    \centering%
    \includegraphics[page=1]{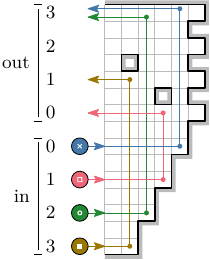}%
    \caption{}%
    \label{fig:function-holes}%
    \vspace{15pt}
    \includegraphics[page=2]{hardness/function-overview}%
    \caption{}%
    \label{fig:function-monotone}%
  \end{subcaptionblock}%
  \hfill%
  \begin{subcaptionblock}{160pt}%
    \centering%
    \includegraphics[page=3]{hardness/function-overview}%
    \caption{}%
    \label{fig:composition-overview}%
  \end{subcaptionblock}%
  \hfill%
  \begin{subcaptionblock}{120pt}%
    \centering%
    \includegraphics[page=4]{hardness/function-overview}%
    \caption{}%
    \label{fig:composition-connections}%
  \end{subcaptionblock}%
  \caption{%
    \capref{fig:function-holes}~Realizing the function~$f = (3,0,3,1)$ in an environment with holes.
    The sequence~\moveseq{ru} transfers the particle from input row~$i$ ($R^{\mathrm{in}}_i$ in the text) to output row~$f(i)$ ($R^{\mathrm{out}}_{f(i)}$).
    \capref{fig:function-monotone}~Holes can be eliminated for monotone functions, such as $(0,0,0,1)$, by extending them to the boundary on the right (shaded area).
    \capref{fig:composition-overview}~Compositions of two functions $f$ and~$g$ are made possible by connecting input and output rows for element~$i$ to node~$p_i$.
    \moveseq{ruld} and \moveseq{rd} transfer a particle from~$p_i$ to the $i$th input row of~$f$ and~$g$, respectively.
    \moveseq{lu} brings the particle from output row~$j$ to~$p_j$.
    \capref{fig:composition-connections}~The $j$th copy of the construction is connected to a common column to the left at node~$p_{h(j),j}$, its copy of node~$p_{h(j)}$, whereas other $p_{i,j}$ are connected to traps.
    Particles are initially placed at nodes~$p_{j,j}$.
  }%
  \label{fig:function-overview}%
\end{figure}

\paragraph*{Representing function application}

For the sake of concise notation, we identify a function~$f$ over the domain~$[m]$ with the $m$-tuple $(f(0), \ldots, f(m-1))$.
We represent the domain~$[m_{\mathrm{in}}]$ and the codomain~$[m_{\mathrm{out}}]$ of a function~$f$ as two sets of equally-spaced rows enclosed from above and below by the outer boundary: \emph{input~rows}~$R_i^\mathrm{in}$, labeled $i \in [m_{\mathrm{in}}]$ from top to bottom, below \emph{output~rows}~$R_j^\mathrm{out}$, labeled $j \in [m_{\mathrm{out}}]$ from bottom to top.
To~represent the application of~$f$, we consider a single particle initially positioned in a row~$R_i^{\mathrm{in}}$.
A~fixed sequence of moves shall transfer the particle to row~$R_{f(i)}^{\mathrm{out}}$.
This is easily accomplished with holes, see Figure~\ref{fig:function-holes}:
A~staircase pattern extends every $R_i^{\mathrm{in}}$ to a distinct column~$C_i$.
Unless $f(i) = m_{\mathrm{out}} - 1$, a unit-size hole is placed in~$C_i$ above row~$R_{f(i)}^{\mathrm{out}}$.
Thus, the sequence \moveseq{ru} transfers the particle from~$R_i^{\mathrm{in}}$ to~$R_{f(i)}^{\mathrm{out}}$.
Note that this construction can be made hole-free if $f$ is monotone, i.e., if $f(i) \leq f(i+1)$ holds for all~$i \in [m_\mathrm{in}-1]$, by extending every hole to the boundary on the right; see Figure~\ref{fig:function-monotone}.

\paragraph*{Building arbitrary compositions of two functions}

Apart from a second function, we also require a way to connect them that allows for arbitrary composition.
We realize two given functions~$f,g: [m] \to [m]$ by embedding them in a function~$F: [2m] \to [m]$ that maps the $i$th even number to~$f(i)$ and the $i$th odd number to~$g(i)$.
Thus, we can retrieve $f(i) = F(2i)$ and $g(i) = F(2i+1)$.
Composition is implemented with the construction in Figure~\ref{fig:composition-overview}, which is attached to the left of a gadget realizing~$F$.
For every $i \in [m]$ there is a node~$p_i$.
The move sequence~\moveseq{ruld} transfers the particle from~$p_i$ to~$R_{2i}^{\mathrm{in}}$, whereas \moveseq{rd} transfers it to~$R_{2i+1}^{\mathrm{in}}$, the $i$th input row for $f$ and $g$, respectively.
If the particle is in row~$R_j^{\mathrm{out}}$, \moveseq{lu} returns it to~$p_j$.
Assuming $w$ to be a sequence of moves transferring the particle from~$R_i^{\mathrm{in}}$ to~$R_{F(i)}^{\mathrm{out}}$ inside the function gadget ($w = \moveseq{ru}$ for the gadget with holes described above), we define move sequences $\overline{f} \coloneq \moveseq{ruld} w \moveseq{lu}$ and $\overline{g} \coloneq \moveseq{rd} w \moveseq{lu}$.
We extend this notation to compositions by defining $\overline{f_{\ell} \circ \dots \circ f_0} \coloneq \overline{f_0} \ldots \overline{f_{\ell}}$, for $f_0, \ldots, f_{\ell} \in \{f,g\}$.
Then, by construction, $\delta^{\FT}_{\overline{f'}}$ turns the configuration that assigns label~$d$ to~$p_i$ and $\bot$ to all~$q \neq p_i$ into one that maps $p_{f'(i)}$ to~$d$ and all other nodes to~$\bot$, for any composition $f' \in \{f,g\}^*$.

\paragraph*{Checking if a given function is generated}

The construction so far allows us to simulate arbitrary compositions of~$f$ and~$g$ applied to a fixed $i \in [m]$ by placing a single particle at~$p_i$ and moving it via~$\overline{f}$ and~$\overline{g}$.
To simulate the effect of a composition on \emph{all} elements $i \in [m]$ at the same time, we stack $m$ copies of this construction.
Let $p_{i,j}$ denote the node~$p_i$ in copy $j$.
The initial configuration places a particle at all nodes~$p_{j,j}$.
We need it to be possible to bring all particles to the leftmost column if and only if a given function~$h$ is generated by~$f$ and~$g$.
To this end, we modify the combined construction to the left of~$p_{i,j}$, depending on~$i$ and $j$, see Figure~\ref{fig:composition-connections}.
Nodes~$p_{h(j),j}$ get their row segment extended to the left into a common column.
Next to all other $p_{i,j}$, $i \neq h(j)$, we place a \textsf{T}-shaped \newterm{trap}.
Once a particle enters a trap and moves up or down, the trap will always contain a particle, rendering it impossible to bring all of them to the leftmost column.

\begin{figure}[tb]
  \centering%
  \begin{subcaptionblock}{150pt}%
    \centering%
    \includegraphics[page=1]{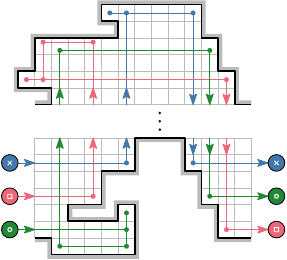}%
    \caption{}%
    \label{fig:adjacent-transposition}%
  \end{subcaptionblock}%
  \hfill%
  \begin{subcaptionblock}{230pt}%
    \centering%
    \includegraphics[page=2,width=220pt]{hardness/adjacent-transposition}%
    \caption{}%
    \label{fig:function-example-simple}%
  \end{subcaptionblock}%
  \caption{%
    \capref{fig:adjacent-transposition}~Gadget for adjacent transpositions.
    The move sequence \moveseq{ruld lurd} transfers a particle from the middle input row to the bottom one, and vice versa.
    Should the particle start in the top input row, it ends in the same row and is only translated to the right.
    \capref{fig:function-example-simple}~Hole-free function gadget for $f = (3,0,3,1)$, decomposed into $f = f_\leq \circ f_2 \circ f_1$ for parallel adjacent transpositions $f_1 = (1,0,3,2)$ and $f_2 = (0,2,1,3)$, and monotone function $f_\leq = (0,1,3,3)$.
    $(\moveseq{ruld lurd})^2$\moveseq{ru} transfers a particle from~$R_i^{\mathrm{in}}$ to~$R_{f(i)}^{\mathrm{out}}$ before \ldir{} moves it out of the gadget; cf.~Figure~\ref{fig:function-holes}.
  }%
  \label{fig:adjacent-transposition-example}%
\end{figure}

\paragraph*{Function gadgets without holes}

Hole-free function gadgets can be implemented by decomposing a function~$f$ into simpler functions that can be realized by skipping back and forth between two contiguous parts of the boundary.
We have already seen how monotone functions can be realized without holes.
Every function~$f$ over the domain~$[m]$ can be written as a composition $f_\leq \circ f_\pi$, where $f_\pi$ is a permutation that sorts elements by their image under~$f$, and $f_\leq$ is a monotone function that maps the rearranged elements to their image under~$f$.
In~turn, $f_\pi$ can be decomposed into a sequence of $\ell \leq m$ parallel transpositions of disjoint pairs of adjacent elements, as in odd-even transposition sort~\cite{taocp-vol3}.
A~gadget for adjacent transpositions is depicted in Figure~\ref{fig:adjacent-transposition}.
It uses the sequence of moves~$v = \moveseq{ruld lurd}$.
For an adjacent pair $(i,i+1)$, activating $v$ transfers the particle from~$R_i^{\mathrm{in}}$ to~$R_{i+1}^{\mathrm{in}}$, and vice versa, while translating the particle horizontally.
If an input row does not participate in a transposition with one of its neighbors, the particle is translated to the right in the same row.
By chaining $\ell$ of these gadgets culminating in a monotone function gadget for $f_\leq$, we obtain a hole-free function gadget for~$f$ with move sequence $w' = v^\ell \moveseq{ru}$; see Figure~\ref{fig:function-example-simple}.
Thus, we can replace the previous non-simple function gadgets with simple ones and set $\overline{f} = \moveseq{ruld}w' \moveseq{lu}$ and $\overline{g} = \moveseq{rd}w' \moveseq{lu}$.

\begin{figure}[tb]
  \centering%
  \begin{subcaptionblock}{270pt}%
    \centering%
    \includegraphics[page=1]{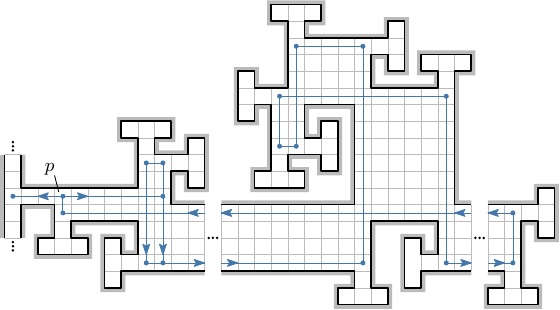}%
    \caption{}%
    \label{fig:forcing}%
  \end{subcaptionblock}%
  \hfill%
  \begin{subcaptionblock}{85pt}%
    \centering%
    \includegraphics[page=2]{hardness/forcing}%
    \caption{}%
    \label{fig:reduction-overview}%
  \end{subcaptionblock}%
  \caption{%
    \capref{fig:forcing}~Gadget enforcing legal move sequences.
    The middle part is repeated $\ell \geq 0$ times.
    Any move sequence alternating between horizontal and vertical directions that transfers a particle initially positioned at~$p$ to the common column must have as prefix one of~\ldir{}, $\overline{f}$, or $\overline{g}$ -- or the particle would get trapped.
    \capref{fig:reduction-overview}~Schematic overview of the full construction.
    For every $i \in [m]$ a gadget~$B_i$ realizing compositions of~$f$ and~$g$ applied to~$i$ is attached to a common leftmost column at the node representing~$h(i)$.
    At the bottom, the forcing gadget~$B_{\mathrm{force}}$ is attached to the same column.
  }%
  \label{fig:forcing-overview}%
\end{figure}

\paragraph*{Enforcing the required move sequences}

The non-simple function gadgets have the advantage that deviations from their move sequences are permissible.
Sequences moving the particles to the leftmost column always correspond to decompositions of~$h$ into~$f$ and~$g$.
This is no longer the case for the hole-free gadgets for adjacent transpositions, where deviating from $\moveseq{ruld~lurd}$ can lead to illegal solutions.
We introduce a final gadget to enforce the required sequences, see Figure~\ref{fig:forcing}.
The left of three parts is connected to the common leftmost column and contains an initially occupied node~$p$.
A~middle part is repeated $\ell$~times and attached to the concluding third part to the right.
Note that, due to the idempotence and regularity of the four basic full-tilt moves, any sequence of moves can be turned into an equivalent one that alternates between horizontal and vertical directions.
The fundamental idea of the gadget is that every such sequence leads the particle starting at~$p$ into a trap, except sequences with prefixes $\overline{f}$, $\overline{g}$, or \ldir{}, which correspond to applying~$f$ or~$g$, or checking if $h$ has been generated, respectively.
This is easily verified by inspection.
The idea to enforce a limited number of possible sequences by having all other sequences trap a particle has been inspired by Friedman's work~\cite{friedman-diss}.

Finally, we can sketch the proof of this section's main theorem, which we re-state here.

\restateHardnessMainThm*
\begin{proof}[Proof sketch]

\begin{figure}[tbp]
  \centering%
  \includegraphics[width=\textwidth]{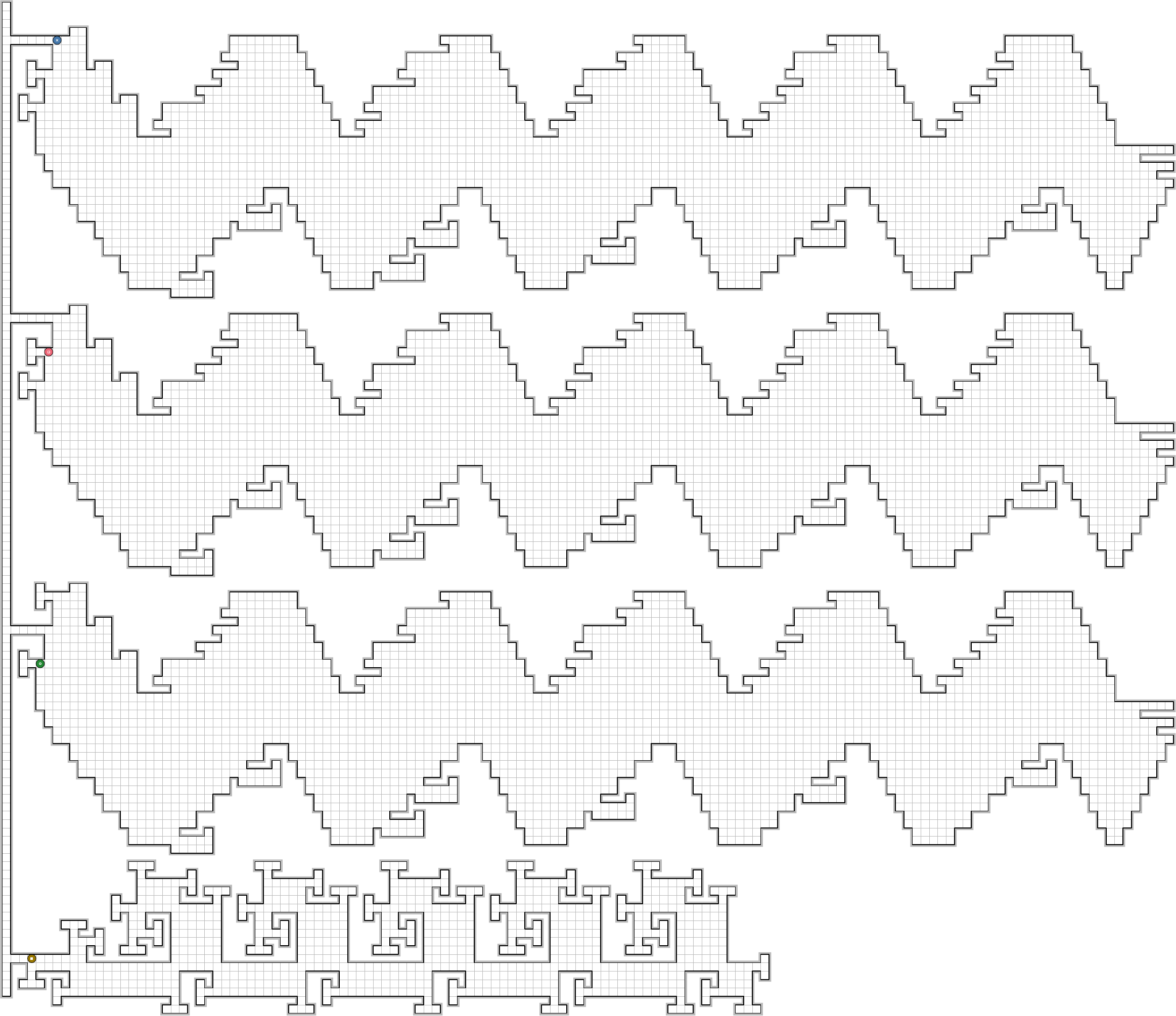}%
  \caption{%
    Complete example of the reduction for given \probname{FFG-2} instance $f = (2,1,2)$, $g = (1,2,0)$, and $h = (0,0,1)$.
    This is a ``yes''-instance, with $h = g^2 \circ f \circ g^2 \circ f \circ g^2$.
    The corresponding sequence of moves is $(\moveseq{rd}v^5 \moveseq{rulu})^2 \moveseq{ ruld}v^5 \moveseq{rulu } (\moveseq{rd}v^5 \moveseq{rulu})^2 \moveseq{ ruld}v^5 \moveseq{rulu } (\moveseq{rd}v^5 \moveseq{rulu})^2 \moveseq{ l}$, where $v = \moveseq{ruld lurd}$, consisting of $373$ moves in total.
  }%
  \label{fig:hardness-instance}%
\end{figure}

Given an instance $f,g,h: [m] \to [m]$ of \probname{FFG-2}, we construct a simple environment and configuration as detailed above.
See Figure~\ref{fig:reduction-overview} for a schematic overview and Figure~\ref{fig:hardness-instance} for an example arising from a specific instance.
The resulting environment is rather large, comprising $\bigO(m^4)$ nodes, but easily computed in polynomial time.

If $h = f_d \circ \cdots \circ f_1$ for some $f_1, \ldots, f_d \in \{f,g\}$, then, by induction on~$d$, $\overline{h} = \overline{f_d} \circ \cdots \circ \overline{f_1}$ transfers the particle starting at~$p_{i,j}$ to~$p_{h(i),j}$, while the particle from~$p$ returns to~$p$.
Thus, $\overline{h}\ldir$ transfers all particles to the leftmost column.

Conversely, every sequence of moves $w \in \D^*$ transferring all particles to the leftmost column must move the particle from~$p$ there.
We argue that the maximal prefix of~$w$ preceding the moment that particle enters the leftmost column must be equivalent to a sequence $w' \coloneq w_d \circ \cdots \circ w_1$, with $w_{\alpha} \in \{\overline{f}, \overline{g}\}$ (since the particle does not get trapped and we may assume the sequence to alternate between horizontal and vertical directions).
Thus, every other particle is at a node $p_{h'(i),j}$ at that time, where $h'$ is the composition of~$f$ and~$g$ corresponding to $w'$.
Since the \ldir{}-move required to bring a particle to the leftmost column does not trap any particle, we can conclude that $h' = h$, i.e., $h$ is generated by $\{f,g\}$.
\end{proof}

It is time to prove \PSPACE-hardness of the fundamental problems of interest.
We define a \newterm{dead end of length $\ell$} to be a contiguous subset of a segment that contains $\ell$ nodes, which include one end of the segment and only nodes that have no neighbors outside the segment.

\begin{restatable}{lemma}{restateDeadEndFT}\label{lem:dead-end-ft}
Let $B=(V,E)$ be an environment, $S \subseteq V$ a column segment that has dead ends of length~$\ell$ at both ends, for some $\ell \in \N$, $p$ the $\ell$th node from the top in~$S$, and $C$ a configuration of~$B$ with no occupied node in~$S$.
Then it holds in $\FT$ that it is possible to occupy $p$ if and only if it is possible to occupy $\ell$ nodes of~$S$ at once.
\end{restatable}

\begin{corollary}\label{cor:hardness-occupancy-ft}
\reconfig{}, \probname{Relocation}, and \occupancy{} are \PSPACE-complete for simple environments in $\FT$.
\end{corollary}
\begin{proof}
The problems are known to be in \PSPACE{} for general environments~\cite{tilt-soda20}.
To show \PSPACE-hardness, we extend the construction from Theorem~\ref{thm:hardness-main-thm} by adding dead ends of length $m+1$ to the top and bottom of the common column, see Figure~\ref{fig:reduction-overview}.
For \occupancy{}, we ask if it is possible to occupy the node~$q$ at position $m+1$ from the top in the common column; \PSPACE-hardness follows by combining Theorem~\ref{thm:hardness-main-thm} and Lemma~\ref{lem:dead-end-ft}.
The argument extends to \probname{Relocation} by giving the particle in~$B_{\mathrm{force}}$ a unique label and asking if a particle with that label can occupy $q$, and to \reconfig{} by asking if it is possible to occupy exactly the topmost $m+1$ nodes in the common column.
\end{proof}

\paragraph*{A note on constructibility}

Contrary to the positive results from Section~\ref{sec:permutations}, the construction from Theorem~\ref{thm:hardness-main-thm} is not constructible by dropping individual assembly particles onto a seed.
The main impediment is the forcing gadget with its many traps.
We claim, without proof, that the environment is still constructible on layer~$d$ of $\mathcal{D}$~\cite{tilt-soda20}, for some constant~$d$, and leave it as an open question if the problems remain \PSPACE-hard for environments constructible on the first layer of~$\mathcal{D}$.

%% file: include/04-2-vacancy.tex
\subsection{PSPACE-completeness of Vacancy}
\label{sec:vacancy}

\begin{figure}[tb]
  \centering%
  \includegraphics{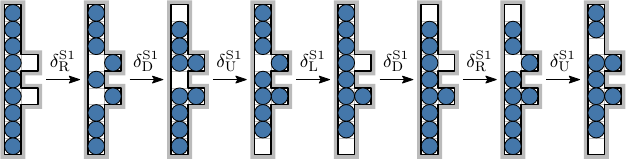}%
  \caption{%
    Contrary to~$\FT$, $\SSt$ does not require~$\ell$ free nodes to free the $\ell$th node from the top in an initially fully occupied column segment with dead ends of length~$\ell$ at both ends. (Here: $\ell=3$.)
  }%
  \label{fig:vacancy-s1-issue}%
\end{figure}

Another open problem concerns the complexity of \vacancy{}: Given an environment~$B$, a configuration~$C$ of~$B$, and a node~$p$, is there a sequence~of~moves~$w$ such that $\delta_w(C)(p) = \bot$?
\vacancy{} is conjectured to be \PSPACE-complete in $\SSt$ by the authors of~\cite{tilt-fill-icalp}, who suggest using ideas from~\cite{tilt-hardness-cccg20} to prove it.
When the authors of~\cite{tilt-hardness-cccg20} showed \PSPACE-hardness of \probname{Relocation} in $\SSt$, they used the observation that $\FT$ and $\SSt$ behave identically when every segment has at most one free node.
Thus, by taking the reduction from Theorem~\ref{thm:hardness-main-thm} and switching occupied and free nodes, we get that it is \PSPACE-hard to decide if a given configuration can be transformed in such a way that the only \emph{free} nodes are in the leftmost column, in both $\FT$ and $\SSt$.
\PSPACE-hardness of \vacancy{} in~$\FT$ easily follows by inverting occupied and free nodes in Lemma~\ref{lem:dead-end-ft}.
It is tempting to think that a similar argument might show the validity of the same reduction for \vacancy{} in~$\SSt$.
However, the analogue of Lemma~\ref{lem:dead-end-ft} for free nodes fails in~$\SSt$, as illustrated in Figure~\ref{fig:vacancy-s1-issue}.

The fix looks easy: Doubling the length of the dead end at one end of the leftmost column is enough.
The high-level idea is that the first contiguous block of occupied nodes in a dead end has a guaranteed minimum size, even in $\SSt$, as long as the dead end is large enough compared to the number of free nodes.
However, a rigorous proof exceeds the space limit of this extended abstract; it builds on a framework from~\cite{tilt-fill-icalp} and can be found in Appendix~\ref{app:vacancy}.

\begin{restatable}{theorem}{restateVacancyHardSSt}\label{thm:vacancy-hard-sst}
\vacancy{} is \PSPACE-complete for simple environments in~$\SSt$.
\end{restatable}

%% file: include/05-conclusions.tex
\section{Conclusions}
\label{sec:conclusions}

A~couple questions remain regarding matrix permutations in simple environments.
Can the $\bigO(\log c)$ upper bound on the sequence length for a fixed permutation be improved, or is there a matching lower bound?
Most of our constructions use the flattening gadget to convert to $(1 \times k)$ rectangles, which increases the space requirement for the bounding box.
Is there a more space-efficient way, maybe one that works directly on $(w \times h)$ rectangles?

We have used matrix permutations as a first example of goals that are achievable in environments constructible by tilt assembly.
While these turned out to be realizable on the first level of the drop shape hierarchy, this may be unattainable for more complicated tasks.
A~general approach for future work may be to make use of the whole hierarchy and compose constructions for fundamental tasks to solve more ambitious challenges.

For example, consider the simulation of bounded universal Turing machines from~\cite{tilt-deterministic}.
The authors show that it is not only possible to apply a TM to a given input but also to write the input onto the tape using tilt moves.
However, the environments realizing the TMs are not constructible by tilt assembly, because they use carefully placed holes as obstacles.
Is it possible to start one step earlier and construct environments capable of computation before programming them and finally running the program, all within the tilt paradigm?

%% file: include/0A-appendix.tex
\section{Appendix}\label{sec:appendix}

\input{include/0A-3-permutations}
\input{include/0A-4-complexity}

%% file: include/0A-3-permutations.tex
\subsection{Omitted details from Section~\ref{sec:permutations-fixed}}

\restateFixedFlatPerm*
\begin{proof}
  Consider a (non-matrix) permutation~$\pi$ of $[k]$.
  To sort~$\pi$ via radix sort, elements are recursively partitioned bitwise, starting with the least-significant bit.
  In each round, relative order is maintained within the partitions.
  After $\lceil \log k \rceil$ partitioning rounds, the permutation is sorted~\cite{taocp-vol3}.
  This approach can be used to reconfigure between arbitrary pairs of permutations of the same integers via relabeling.

  To emulate a single partitioning round, we introduce a subset partitioning gadget, see Figure~\ref{fig:fixed-labeled-subset-picker-a} on page~\pageref{fig:fixed-labeled-subset-picker-a}.
  The particles enter the gadget from the top left.
  After move sequence $\moveseq{dru}$, particles from the chosen subset (in this case, odd labels) are located in elevated nooks with no room for horizontal movement, and after an $\ldir$ move, they are separated from the rest of the particles without changing the relative order within either subset, see Figure~\ref{fig:fixed-labeled-subset-picker-b} on page~\pageref{fig:fixed-labeled-subset-picker-b}.
  With moves $(\moveseq{drul})^2$, the two subsets get concatenated and exit the gadget.

  By chaining $\lceil \log k \rceil$ of the gadgets, one for each position in the labels' bit strings, we can emulate radix sort to realize any fixed $(k \times 1)$-matrix permutation in $(\moveseq{drul})^{3 \lceil \log k \rceil + 2} = \bigO(\log k)$ steps, see Figure~\ref{fig:fixed-labeled-permutation} on page~\pageref{fig:fixed-labeled-permutation}.
  The connector and partitioning gadgets are designed to enable pipelining multiple particle configurations simultaneously with repeated moves $\moveseq{drul}$.

  Each partitioning gadget has width $5k + 1$ and height~$6$ (excluding entry and exit).
  These gadgets are stacked on top of each other with constant offset and connected via a connector of width $2k + 1$ and height $7 \lceil \log k \rceil - 2$, resulting in a bounding box of dimensions $\bigO(k) \times \bigO(\log k)$.
\end{proof}

\restateFixedKPerm*
\begin{proof}
  Without loss of generality, let $h \leq w$.
  Otherwise, rotate the instance.
  To convert $(w \times h)$-matrices to $(k \times 1)$-matrices, we use a flattening gadget, see Figure~\ref{fig:matrix-flattener} on page~\ref{fig:matrix-flattener}.
  A $(w \times h)$-matrix enters the gadget from the left and gets converted into a $(k \times 1)$-matrix with moves $(\ddir)\moveseq{rul}$.
  It is then moved to the $(k \times 1)$ construction from Theorem~\ref{thm:fixed-flat-k-perm} with moves $(\moveseq{drul})^2$.
  Once it returns from this construction from the right, it is turned back into a $(w \times h)$-matrix and transported to the exit of the flattening gadget with moves $(\moveseq{drul})^2$.

  The flattening gadget has a width of $2w + 3k + 1$ and a height of $4h + 6$.
  Therefore, it fits in a bounding box of dimensions $\bigO(w + k) \times \bigO(h) = \bigO(k) \times \bigO(h)$ under the assumption $h \leq w$.
  For general $w$ and $h$, this corresponds to $\bigO(k) \times \bigO(\min \{w, h\})$.
  Combined with the construction from Theorem~\ref{thm:fixed-flat-k-perm}, this yields the stated result.
\end{proof}

\restateFixedCPerm*
\begin{proof}
  We only consider $(k \times 1)$-matrices.
  For other dimensions, we can reuse the flattening gadget introduced in the proof of Corollary~\ref{cor:fixed-k-perm}, see Figure~\ref{fig:matrix-flattener} on page~\pageref{fig:matrix-flattener}.

  To invert a subset partitioning step, we construct a gadget that can individually interleave the $q$ rightmost particles of a $(k \times 1)$-matrix between the $k - q$ leftmost particles without changing their relative order.
  The gadget consists of two stages.
  First, the $k - q$ leftmost particles are reversed, see Figure~\ref{fig:inverse-subset-picker-invert} on page~\pageref{fig:inverse-subset-picker-invert}.
  Then, the~$q$ rightmost particles are interleaved between the $k - q$ leftmost particles, which also restores the original relative order of these particles, see Figure~\ref{fig:inverse-subset-picker-insert}.
  Both subgadgets perform the desired operation with move sequence $(\moveseq{drul})^3$.

  Now, for both~$M_s$ and~$M_t$, we compute sorting sequences with $\lceil \log c \rceil$ partitioning steps via radix sort.
  We first chain the $\lceil \log c \rceil$ corresponding subset partitioning gadgets to sort~$M_s$ as done in Theorem~\ref{thm:fixed-flat-k-perm}.
  Then, we invert the partitioning sequence computed for sorting~$M_t$ and chain the corresponding inverse subset partitioning gadgets.
  Combined, this allows realizing the matrix permutation from~$M_s$ to~$M_t$ in $(\moveseq{drul})^{9\lceil \log c \rceil + 2} = \bigO(\log c)$ steps.

  The subgadgets for the first and second stage of the inverse gadget, excluding entry and exit, have width at most~$4k - 1$ and $3k - 1$, and height at most $k + 4$ and $4k + 3$, respectively.
  Therefore, the combined gadget fits in a bounding box of size $\bigO(k) \times \bigO(k)$.
  By chaining the $2\lceil \log c \rceil$ subset partitioning and inverse subset partitioning gadgets vertically, the entire construction fits in a bounding box of size $\bigO(k) \times \bigO(k \log c)$.
\end{proof}

\subsection{Omitted details from Section~\ref{sec:permutations-arbitrary}}

\restateArbitraryFast*
\begin{proof}
  As before, we only consider $(k \times 1)$-matrices and reuse the flattening gadget of Corollary~\ref{cor:fixed-k-perm} (see Figure~\ref{fig:matrix-flattener} on page~\pageref{fig:matrix-flattener}) for other dimensions.

  We construct an environment that allows splitting the particles into two ``stacks'' in $\bigO(k)$ steps, retaining the relative order of particles within the stacks.
  The particles can be repeatedly popped from either of the two stacks in $\bigO(k)$ total steps.
  These operations allow us to emulate the $\bigO(\log c)$-move radix-based strategy for fixed permutations from Theorem~\ref{thm:fixed-c-perm} in $\bigO(k \log c)$ steps for any compatible permutation.
  To emulate subset partitioning, move all particles belonging to the desired subset to the left stack and the remaining particles to the right stack, and subsequently empty one stack after the other.
  To invert the same operation, split the particles into the two stacks again, but pop them in the inverse push-order.

  The full construction is shown for six particles in Figure~\ref{fig:two-stacks} on page~\pageref{fig:two-stacks}.
  A round is initiated with moves~$\rdir \ddir$.
  Afterward, a particle is moved to the left stack with $\moveseq{ldrd}$ and to the right stack with $\moveseq{rd}$.
  Once all particles have been pushed, the left stack can be popped with $\moveseq{rurdlulurd}$ and the right stack with $\moveseq{lulurd}$.
  Finally, the particles are prepared for the next round with $\moveseq{ld}$.
  In total, each round consists of up to $4 + 4k + 10k = \bigO(k)$ moves and any compatible permutation can be realized within $\bigO(\log c)$ rounds.
  The $(k \times 1)$-construction has width $4k + 1$ and height~$8$.
  Together with the flattening gadget of Corollary~\ref{cor:fixed-k-perm}, the full construction fits in a bounding box of dimensions $\bigO(k) \times \bigO(\min \{w, h\})$.
\end{proof}

\restateBubbleSort*
\begin{proof}
  Assume $h \leq w$.
  Otherwise, rotate the instance.
  We construct an environment that consists of two gadgets for basic operations: \emph{swap} and \emph{shift}.
  The first operation swaps the particle in the top-left corner with its right neighbor.
  The second shifts all inner particles one position to the left and all leftmost particles to the right position of the previous row.
  The particle in the top left corner gets moved to the bottom right corner of the matrix.
  With these two operations, we can emulate bubble sort:
  If the first two elements are in the wrong order, we swap.
  Then, we shift once to emulate increasing the index and repeat.
  After~$k$ shifts, the index is back at the starting position.
  We have to repeat the whole procedure up to~$k$ to reach any desired target permutation with a total of $\bigO(k^2)$ steps.

  \begin{figure}
    \centering%
    \includegraphics{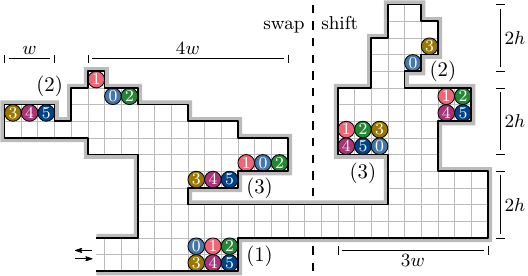}%
    \caption{%
      The bubble sort construction for arbitrary matrix permutations from Theorem~\ref{thm:arbitrary-c-perm-fast} with $k = 6$ particles.
      The particles start at the entry / exit of the environment in sorted order (1).
      For the swap operation (left half), they are moved to the entry of the swap gadget with moves $\moveseq{ul}$ and the topmost row is separated from the rest of the matrix with moves $\moveseq{ulur}$ (2).
      Here, the particle at the second position of the topmost row is fixed to move it to the front.
      Then, with moves $\moveseq{drdr}$, the two halves of the matrix are recombined (3).
      Finally, the particles are moved back to their starting position with moves $\moveseq{ld}$.
      For the shift operation (right half), the particles are moved from the start to the shift gadget entry with moves $\moveseq{urul}$ and the leftmost column of the matrix is separated and shifted with moves $\moveseq{urd}$ (2).
      Then, the separated and shifted column is moved to the right of the remaining matrix with moves $\moveseq{ld}$ (3), which corresponds to the shift operation described in the proof of Theorem~\ref{thm:arbitrary-c-perm-fast}.
      Finally, the particles are moved back to their starting position with moves $\moveseq{rdldr}$.
    }%
    \label{fig:bubble-sort}%
  \end{figure}

  The environment is shown in Figure~\ref{fig:bubble-sort}.
  If the matrix is at the entry or exit of the environment, we can perform a swap operation and return with moves $\moveseq{ul ulurdrdldr}$, regardless of the matrix dimensions.
  Similarly, a shift operation is done via moves $\moveseq{urul urdldrdldr}$.
  The environment has width $8w + 5$ and height $7h + 2$ and therefore fits in a bounding box of size $\bigO(w) \times \bigO(h)$.
\end{proof}

\restateLowerBound*
\begin{proof}
  For simplicity, let $c \mid k$ and assume that each color occurs $k/c$ times in~$M$.
  We divide the~$k$ particle positions into $k/c$ groups where each group needs to contain each color at least once, yielding $(c!)^{k/c}$ compatible matrices.
  While this does not capture all $k!/((k/c)!)^c$ possible compatible matrices, it is a lower bound.
  On the other hand, the number of reachable configurations within~$\ell$ moves is bounded by $4^0 + 4^1 + \cdots + 4^\ell = 1/3 (4^{\ell + 1} - 1) \leq 4^{\ell + 1}$.
  Then\[
    4^{\ell + 1} \geq \sum_{i = 0}^\ell 4^i \geq \frac{k!}{((k/c)!)^c} \geq (c!)^{k/c} \implies \ell + 1 \geq (k/c) \cdot c \log c = k \log c.\qedhere
  \]
\end{proof}

\subsection{Omitted details from Section~\ref{sec:permutations-assembly}}

\begin{figure}
  \begin{subcaptionblock}{197pt}%
    \raggedright%
    \includegraphics[page=2,scale=.875]{assembly/fixed-colored-permutation}%
    \caption{}%
    \label{fig:assembly-flattening}%
  \end{subcaptionblock}%
  \hfill%
  \begin{subcaptionblock}{141pt}%
    \raggedleft%
    \includegraphics[page=5,scale=.875]{assembly/fixed-colored-permutation}%
    \caption{}%
    \label{fig:assembly-interleave}%
  \end{subcaptionblock}%
  \medskip

  \begin{subcaptionblock}{190pt}%
    \raggedright%
    \includegraphics[page=3,scale=.875]{assembly/fixed-colored-permutation}%
    \caption{}%
    \label{fig:assembly-subset}%
  \end{subcaptionblock}%
  \hfill%
  \begin{subcaptionblock}{190pt}%
    \raggedleft%
    \includegraphics[page=4,scale=.875]{assembly/fixed-colored-permutation}%
    \caption{}%
    \label{fig:assembly-inverse}%
  \end{subcaptionblock}%
  \caption{%
    Assembly order for the gadgets in the fixed-permutation environments.
    Green assembly particles are dropped from the top, blue particles from the left, yellow particles from the bottom, and red particles from the right.
    The gadgets are attached to gray assembly particles from previous construction steps.
    \capref{fig:assembly-flattening}~The flattening gadget from Figure~\ref{fig:matrix-flattener}.
    \capref{fig:assembly-interleave}~The second stage of the inverse subset partitioning gadget from Figure~\ref{fig:inverse-subset-picker-insert}.
    \capref{fig:assembly-subset}~The subset partitioning gadget from Figure~\ref{fig:fixed-labeled-subset-picker}.
    \capref{fig:assembly-inverse}~The first stage of the inverse subset partitioning gadget~from~Figure~\ref{fig:inverse-subset-picker-invert}.
  }%
\end{figure}

\restatePermutationAssembly*
\begin{proof}
  We begin with the two-stack construction from Theorem~\ref{thm:arbitrary-c-perm-fast}, see Figure~\ref{fig:two-stacks}, as it is the smallest and simplest environment.
  The assembly shape and assembly order for the $k \times 1$ construction with $k = 6$ is shown in Figure~\ref{fig:assemble-two-stacks} on page~\pageref{fig:assemble-two-stacks}.
  To assemble the construction for larger values of~$k$, it suffices to cut the assembly shape vertically along the four dashed lines in Figure~\ref{fig:assemble-two-stacks} and insert assembly particles in the resulting gaps (e.g., insert $k - 6$ particles from the right in the gap between $26$ and $27$).
  Analogously, remove assembly particles adjacent to the dashed lines for constructions with $k < 6$.
  Counting the number of assembly particles reveals that the obstacle shape consists of $13k + 22 = \bigO(k)$ obstacles.

  To extend the construction to other dimensions, assemble the flattening gadget from the proof of Corollary~\ref{cor:fixed-k-perm} at the entry of the two-stack construction.
  See Figure~\ref{fig:assembly-flattening} for turning a $3 \times 2$-matrix into a $6 \times 1$-matrix.
  The gadget is assembled to the left of the construction with only moves $\ddir$, $\rdir$, and $\udir$ so there is no interference with previously dropped particles.
  The same idea works for other dimensions, by splitting the gadget appropriately and repeating the corresponding drop patterns in the gaps.
  The flattening gadget can be constructed with $9wh + 3w + 8h + 13 = \bigO(k)$ assembly particles.

  The assembly shape and order for the $(3 \times 2)$ bubble sort environment from Figure~\ref{fig:bubble-sort} (for Theorem~\ref{thm:arbitrary-c-perm-bubble}) is shown in Figure~\ref{fig:assemble-bubble-sort}.
  As before, the dashed lines show where to cut the shape (horizontally and vertically) and insert or remove particles for other matrix dimensions.
  As there is a constant number of segments in the obstacle shape and their length is always bounded by $\bigO(w)$ or $\bigO(h)$, the total number of obstacles is $\bigO(w + h)$.

  \begin{figure}
    \centering%
    \includegraphics{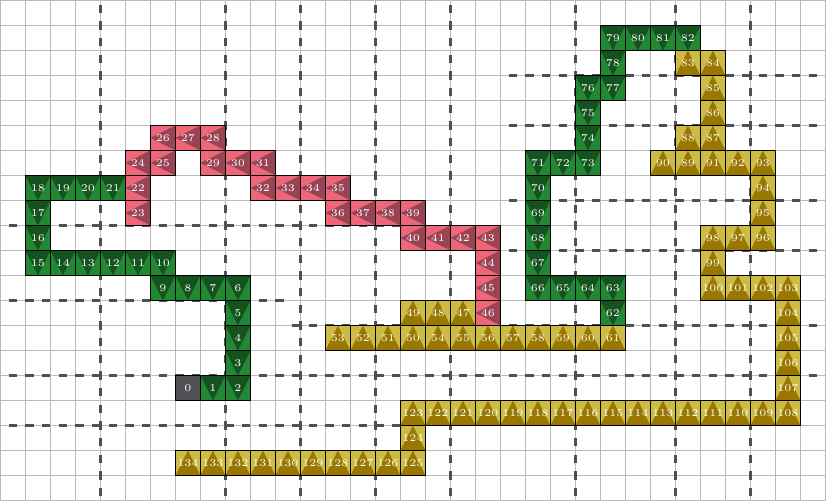}%
    \caption{%
      Assembly order for the bubble sort construction from Figure~\ref{fig:bubble-sort} with seed particle~$0$.
      Green assembly particles are dropped from the top (1--22 and 63--83), red particles from the right (23--47), and yellow particles from the bottom (48--62 and 84--135).
      The dashed lines show where to add or remove assembly particles to extend the construction to other matrix dimensions.
    }%
    \label{fig:assemble-bubble-sort}%
  \end{figure}

  It remains to examine the constructibility of the environment for fixed $c$-colored $(w \times h)$-matrix permutations from Theorem~\ref{thm:fixed-c-perm}, which consists of a sequence of gadgets.
  As we have already examined the matrix flattening gadget, we focus on the remaining (inverse) subset partitioning gadgets for $(k \times 1)$-matrices.
  Assembly orders for all three stages are shown in Figures~\ref{fig:assembly-interleave}--\subref{fig:assembly-inverse} for $k = 6$.
  Although the exact obstacle shape depends on the chosen subset for the partitioning, the same pattern can be applied for all variations of the gadgets as well as all larger dimensions:
  drop assembly particles from the right for the bottom part, from the top for the right part, and from the left for the top part.
  These movement directions ensure that newly dropped assembly particles do not collide with placed particles when the full construction is built from the bottom up.
  All three gadget constructions clearly use $\bigO(k)$ assembly particles.
  There are $\bigO(\log c)$ copies of these gadgets stacked on top of each other and connected via a simple tunnel, resulting in $\bigO(k \log c)$ obstacles for the full obstacle shape.
  The full drop shape for a $4$-colored $(3 \times 2)$-matrix is shown in Figure~\ref{fig:assemble-fixed-colored}.
  \begin{figure}
    \centering%
    \includegraphics[page=1]{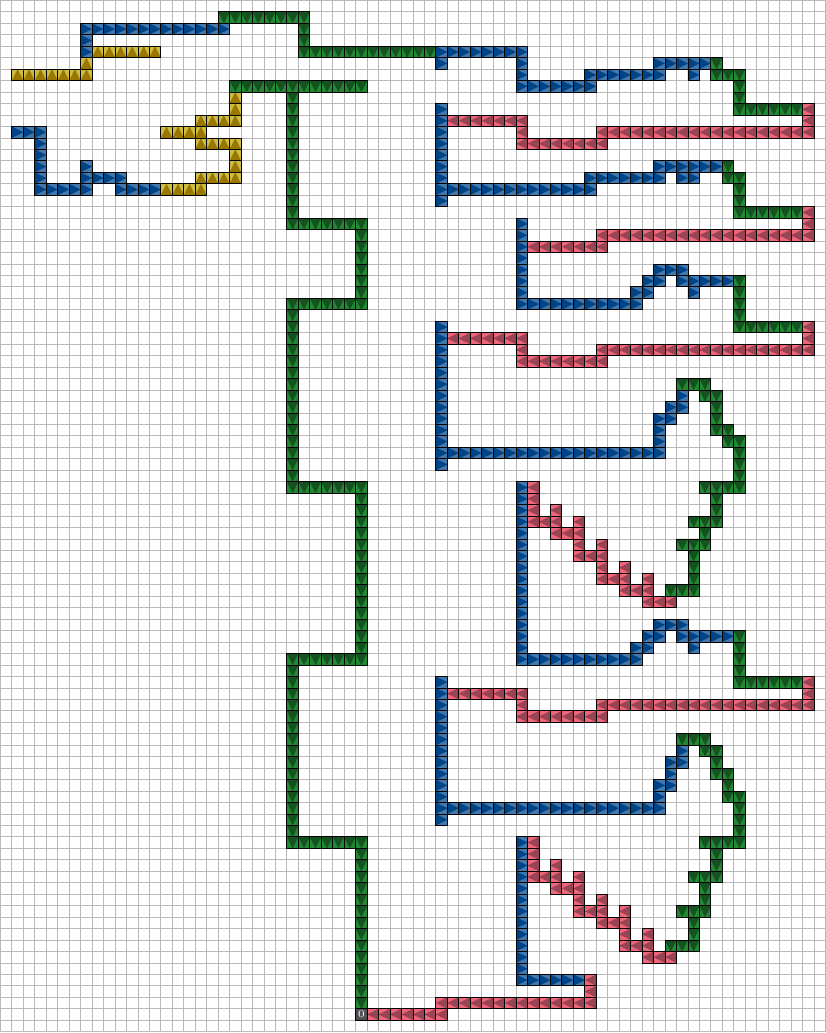}%
    \caption{%
      Assembly order for a $4$-colored $(3 \times 2)$-matrix permutation construction from Theorem~\ref{thm:fixed-c-perm} with seed particle~$0$ at the bottom.
      Red assembly particles are dropped from the right, blue particles from the left, green particles from the top, and yellow particles from the bottom.
      We leave the precise assembly order as well as the permutation realized by this construction for input matrix $(\begin{smallmatrix}1 & 0 & 0 \\ 3 & 3 & 2\end{smallmatrix})$ as recreational exercises to the reader.
      Hint:~The right half of the construction is assembled before the left half.
    }%
    \label{fig:assemble-fixed-colored}%
  \end{figure}
\end{proof}

%% file: include/0A-4-complexity.tex
\subsection{Omitted details from Section~\ref{sec:complexity-simple}}

The following proof of Lemma~\ref{lem:ffg2-hard} uses a transparent construction to reduce an arbitrary number of transformations to exactly two transformations that generate a closely related transformation monoid.
We have seen it used by Vorel~\cite{vorel}, who in turn credits Berlinkov~\cite{berlinkov}, but it may well be folklore.

\restateFFGhard*
\begin{proof}
\probname{FFG-2} is clearly contained in \PSPACE{} as a special case of \probname{FFG}, which is \PSPACE-complete~\cite{kozen}.
To demonstrate \PSPACE-hardness, we give a polynomial-time reduction from \probname{FFG} to \probname{FFG-2}.
Let $S$ be a finite set, $f_i: S \to S$ a family of $j$ functions, $0 \leq i < j$, and $h: S \to S$ another function, as in the input to \probname{FFG}.
We define $S' \coloneq S \times [j]$ and three functions $f, g, h': S' \to S'$ as follows.
\begin{align*}
  f((s, i)) & \coloneq (f_i(s), i)\\
  g((s, i)) & \coloneq (s, (i+1) \bmod j)\\
  h'((s, i)) & \coloneq (h(s), i)
\end{align*}
These functions constitute an input to \probname{FFG-2} with domain $S'$ and can clearly be computed in $(j|S|)^{\bigO(1)}$ time, i.e., polynomial in the size of the input to \probname{FFG}.%
\footnote{The precise degree of the polynomial depends on immaterial questions regarding the representation of the $f_i$; we only assume that they are given explicitly, requiring $\Omega(|S|)$ space each.}

The intuitive idea is that the first component of a pair from $S \times [j]$ keeps track of the current element, and the second component is used to select the next function $f_i$ to apply.
Successive applications of $g$ switch between the $f_i$.
Afterwards, $f$ applies the selected $f_i$ to the current element.
It remains to show that $h$ is generated by $\{f_i : i \in [j]\}$ if and only if $h'$ is generated by $\{f,g\}$.

First, assume $h = f_{i_\ell} \circ \dots \circ f_{i_1}$ for some $\ell \in \N$ and $i_1, \ldots, i_{\ell} \in [j]$.
Then, by construction, $(g^{(-i_\ell) \bmod j} \circ f \circ g^{i_\ell}) \circ \dots \circ (g^{(-i_1) \bmod j} \circ f \circ g^{i_1}) = h'$, i.e., $h'$ is generated by $\{f,g\}$.

Now, assume $h'$ is generated by $\{f,g\}$.
Let $\ell$ be the number of occurrences of $f$ in the composition.
Then there are $\ell + 1$ powers of $g$ interspersed among them (possibly $g^{0}$, if $f$ is applied several times in succession).
We can assume every power of $g$ to be at most $g^{j-1}$, since $g^{m} = g^{m \bmod j}$ for all $m$.
Thus, there is a sequence $i_0, \ldots, i_\ell \in [j]$ such that $h' = g^{i_\ell} \circ f \circ \dots \circ g^{i_1} \circ f \circ g^{i_0}$.
Let $i'_p = (\sum_{q=0}^p i_q) \bmod j$ for $0 \leq p \leq \ell$.
Note that $i'_\ell = 0$ because $h'$ keeps the second component unchanged.
Now, $h = f_{i'_{\ell-1}} \circ \dots \circ f_{i'_0}$, i.e., $h$ is generated by $\{f_i : i \in [j]\}$.
\end{proof}

\restateDeadEndFT*
\begin{proof}
We first consider the backward direction and assume there is $w' \in \D^*$ such that at least~$\ell$ nodes are occupied in~$\delta^{\FT}_{w'}(C)$.
Then $p$ is occupied in $\delta^{\FT}_{w'\udir}(C)$.

For the forward direction, consider a shortest sequence $w \in \D^*$ such that $p$ is occupied in $\delta^{\FT}_w(C)$.
$w$ cannot be empty, because no node from~$S$ is occupied in~$C$.
The last move of~$w$ must be \udir{} or \ddir{}, because horizontal moves do not change the configuration at~$p$.
For $p$ to be occupied in $\delta^{\FT}_{\udir}(C')$, at least~$\ell$ nodes of~$S$ must be occupied in $C'$, for any configuration $C'$; $\ell + 1$ occupied particles are required for \ddir{}.
Since neither \udir{} nor \ddir{} changes the number of occupied particles in a column segment, $w$ satisfies $|\{p \in S: \delta^{\FT}_{w}(C)(p) \neq \bot\}| \geq \ell$.
\end{proof}

\restateHardnessMainThm*
\begin{proof}
We give a polynomial-time reduction from \probname{FFG-2}, shown \PSPACE-hard in Lemma~\ref{lem:ffg2-hard}.
For a given instance $f,g,h: [m] \to [m]$ of \probname{FFG-2}, we construct a simple environment using the gadgets detailed in Section~\ref{sec:complexity-simple}.
We build a hole-free function gadget for the function $F: [2m] \to [m]$ with $2i \mapsto f(i)$ and $2i + 1 \mapsto g(i)$ by decomposing~$F$ into $\ell \leq 2m$ parallel adjacent transpositions (followed by a single monotone function) and chaining the corresponding gadgets.
The function gadget is attached to the right of the gadget from Figure~\ref{fig:composition-overview} on page~\pageref{fig:composition-overview}, which contains nodes $p_0, \ldots, p_{m-1}$.
We create $m$ copies of this partial construction, labeled $B_0, \ldots, B_{m-1}$, and attach $B_j$ to a common column at $p_{h(j),j}$, its copy of node~$p_{h(j)}$, while its other $p_{i,j}$ get connected to traps.
A~forcing gadget with $\ell$ repetitions of its middle part is also attached to the common column.
The initial configuration~$C$ places particles at the copy of $p_i$ in $B_i$, for all $i \in [m]$, and one particle at the node~$p$ of the forcing gadget.
See Figure~\ref{fig:reduction-overview} on page~\pageref{fig:reduction-overview} for a schematic overview and Figure~\ref{fig:hardness-instance} on page~\pageref{fig:hardness-instance} for an example environment arising from a specific instance.

We use the common $\bigOtilde$-notation in the analysis of the time complexity to disregard $\log$-factors arising from the $\lceil \log m \rceil$ bits required to represent an element~$i \in [m]$.
The most noteworthy computational step is the decomposition of~$F$, which is achieved by odd-even transposition sort~\cite{taocp-vol3} in at most $2m$ rounds of $\bigOtilde(m)$ time each.
Every resulting gadget for parallel adjacent transpositions uses $\bigO(m^2)$ nodes and can be build in $\bigOtilde(m^2)$ time.
Since the whole constructions contains $\bigO(m^2)$ of them, constructing these gadgets alone takes $\bigOtilde(m^4)$ time.
This dominates the time required to build the remaining gadgets and assemble the completed environment.
Thus, the procedure runs in polynomial time.

It remains to show that $h$ is generated by $\{f,g\}$ if and only if there is a sequence $w \in \D^*$ such that $\delta^{\FT}_w(C)$ sees all particles in the leftmost column.
For the forward direction, assume $h = f_d \circ \cdots \circ f_1$ for some $f_1, \ldots, f_d \in \{f,g\}$.
By construction of the gadgets, the sequence of moves~$\overline{f_\alpha}$, $\alpha \leq d$, moves a particle positioned at~$p_{i,j}$ to~$p_{f_\alpha(i),j}$, for all $i \in [m]$ in every~$B_j$, and the particle starting at~$p$ in the forcing gadget back to~$p$.
Thus, by induction on~$d$, $\overline{h}$ moves the particle starting at~$p_{j,j}$ to~$p_{h(j),j}$, for all $j \in [m]$, and the particle in the forcing gadget back to~$p$.
Since all of these nodes are connected to the leftmost column, the sequence $\overline{h}\ldir$ transfers all particles there.

For the backward direction, assume there is a sequence $w \in \D^*$ such that $\delta^{\FT}_w$ brings all particles to the leftmost column.
Without loss of generality, we may assume that $w$ alternates between horizontal and vertical directions.
Let~$w'$ be the longest prefix of~$w$ such that no particle is in the leftmost column in~$\delta^{\FT}_{w'}(C)$, i.e., the particles are kept in their original sub-environments, $B_0, \ldots, B_{m-1}$ and $B_{\mathrm{force}}$, at every point of the application of~$w'$.
Furthermore, $w'\ldir$ has at least one particle enter the leftmost column, because \ldir{} is the only move that can accomplish that for the first time.
Consider the action of~$w'\ldir$ on the particle in~$B_{\mathrm{force}}$.
Split $w'\ldir$ into minimal sub-sequences $w_0,\ldots,w_d$ such that each $\delta^{\FT}_{w_\alpha}$, $\alpha \in [d]$, sees the particle starting at~$p$ return to~$p$ (possibly $d=0$ if the particle does not return to~$p$ during the application of~$w'\ldir$).
Then all $w_\alpha$, $\alpha \in [d]$, start with \rdir{}, since the only way to move the particle to~$p$ is via \udir{} and a subsequent \ldir{} would move it to the leftmost column.
By construction of the forcing gadget, there are only two possible ways to extend the $w_\alpha$ without the particle getting trapped: $\moveseq{ruld}(\moveseq{ruld lurd})^\ell \moveseq{rulu}$ and $\moveseq{rd}(\moveseq{ruld lurd})^\ell \moveseq{rulu}$, i.e., $\overline{f}$ and~$\overline{g}$.
The construction of the function gadgets ensures that a particle at $p_{i,j}$ is moved to~$p_{f(i),j}$ by~$\overline{f}$ and to~$p_{g(i),j}$ by~$\overline{g}$.
Let~$f_\alpha \coloneq f$ if $w_\alpha = \overline{f}$ and $f_\alpha \coloneq g$ if $w_\alpha = \overline{g}$, for $\alpha \in [d]$.
Define $h' = f_{d-1} \circ \cdots \circ f_0$.
Then, by induction on~$d$, the sequence $w_0 \ldots w_{d-1}$ moves the particle starting at~$p_{j,j}$ to~$p_{h'(j),j}$, for all~$j \in [m]$.
The final subsequence, $w_d$, must satisfy two properties: It must move one particle to the common column, and it must be a prefix of~\ldir{}, $\overline{f}$, or $\overline{g}$ to not trap the particle in~$B_{\mathrm{force}}$.
Only $w_d = \ldir$ satisfies both properties, because no prefix of $\overline{f}$ or $\overline{g}$ moves a particle starting at~$p$ or any $p_{i,j}$ to the leftmost column.
Since $w_d$ does not trap any of the particles, it must move all particles to the common column.
Thus, we must have $h'(j) = h(j)$ for all~$j \in [m]$, i.e., $h$ is generated by $\{f,g\}$.
\end{proof}

\subsection{Omitted details from Section~\ref{sec:vacancy}}
\label{app:vacancy}

Using the observation from~\cite{tilt-hardness-cccg20} that $\FT$ and $\SSt$ behave identically when every segment has at most one free node, one can adapt the reduction from Corollary~\ref{cor:hardness-occupancy-ft} to show \PSPACE-hardness of \probname{Relocation} for simple environments in $\SSt$ by inverting free and occupied nodes, giving a specific label~$t$ to the particle at the topmost node in the leftmost column, and assigning labels distinct from~$t$ to all other particles, before asking if the node at position $|V|-k+1$ from the top in the leftmost column can be occupied by a (i.e., the only) particle with label~$t$.
This can only be achieved with $k$ free nodes in the leftmost column; see~\cite{tilt-hardness-cccg20} for the original argument.
It is easy to see that hardness of \reconfig{} transfers as well, but note the contrast to~\occupancy{}, which can be solved in polynomial time for arbitrary environments in~$\SSt$~\cite{tilt-limited-directions-jip20}.

\begin{corollary}
\reconfig{} and \probname{Relocation} are \PSPACE-complete for simple environments in~$\SSt$.
\end{corollary}

As mentioned in Section~\ref{sec:vacancy}, this argument does not immediately extend to \vacancy{} in~$\SSt$.
To develop intuition for the sometimes subtle differences between models and to be able to argue rigorously about them, we extend a framework started by the authors of~\cite{tilt-fill-icalp}.

\paragraph*{Generalized models, duality, and simulation}

It can be useful to imagine free nodes not as empty but as containing a different kind of particle, an \emph{anti-particle}.
We extend the notion of configuration to accommodate this idea: In this subsection, a configuration of an environment~$B=(V,E)$ is a mapping $C: V \to \B \times \N$, where the first component $b \in \B = \{0,1\}$ indicates if the node contains a particle ($b=1$) or an anti-particle ($b=0$), and the second component is the (anti-)particle's label.
One can switch between particles and anti-particles with the transformation $\texttt{comp}(C)$ that negates the first component, i.e., $\texttt{comp}(C)(p) \coloneq (1-b, \ell)$ when $C(p) = (b, \ell)$.
A~\emph{model} is a family~$\delta_i$ of \emph{moves}, i.e., transformations of configurations, indexed by a set~$I$ of \emph{inputs}.
As before, we extend $\delta$ to an infinite family indexed by the set~$I^*$ of sequences of inputs.
To examine the behavior of anti-particles, we define the \emph{dual} of a move~$\delta$ as $\overline{\delta} \coloneq \texttt{comp} \circ \delta \circ \texttt{comp}$.
This definition is compatible with the composition of moves, i.e., $\overline{\delta_i \circ \delta_j} = \overline{\delta_i} \circ \overline{\delta_j}$, which allows us to define the \emph{dual model}~$\overline{\delta}$ of a model~$\delta$ as the family of its dual~moves indexed by the same inputs.
We say that a model~$\delta^0$ with inputs~$I_0$ \emph{simulates} a model~$\delta^1$ with inputs~$I_1$ (in symbols $\delta^0 \succeq \delta^1$) if for every configuration~$C$ and sequence $w \in I_1^*$ there is a sequence $w' \in I_0^*$ such that $\delta^0_{w'}(C) = \delta^1_w(C)$.

\begin{observation}\label{obs:occupancy-vacancy-dual}
For every model~$\delta$, there is a polynomial-time reduction from \occupancy{} in~$\delta$ to \vacancy{} in~$\overline{\delta}$ that keeps the environment unchanged.
\end{observation}
\begin{proof}
Apply \texttt{comp} to the initial configuration.
This takes linear time, keeps the environment, and we have $\delta_w(C)(p) = (1,\ell) \iff \overline{\delta_w}(\texttt{comp}(C))(p) = (0,\ell)$ by the definition of duality.
\end{proof}

\paragraph*{A base model with individual control}

Let us imagine an exceptionally powerful mechanism that allows us to control individual particles by swapping them with adjacent anti-particles.
We define a new model~$\IC$ (for \emph{individual control}) with $\delta^{\IC}_i$ indexed by~$i \in V \times \D$ as follows.
If~$v \in V$ has a left neighbor~$u = (v - (1,0)^{\mathrm{T}}) \in V$ such that $C(v) = (1, \ell_v)$ and $C(u) = (0, \ell_u)$, then $\delta^{\IC}_{(v, \ldir)}(C)(v) = (0, \ell_u)$ and $\delta^{\IC}_{(v, \ldir)}(C)(u) = (1, \ell_v)$, while $\delta^{\IC}_{(v, \ldir)}(C)(p) = C(p)$ for all~$p \notin \{u,v\}$.
Otherwise, $\delta^{\IC}_{(v, \ldir)}$ is the identity.
As usual, other directions~$d \neq \ldir$ are defined analogously and $\delta_i^{\IC}$ is extended to indices~$i \in (V \times \D)^*$.

\begin{figure}[tb]
  \centering%
  \begin{subcaptionblock}{124pt}%
    \centering%
    \includegraphics[page=1]{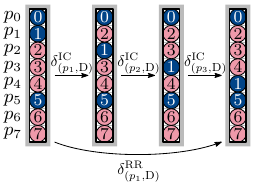}%
    \caption{}%
    \label{fig:models-rr}%
  \end{subcaptionblock}%
  \hfill%
  \begin{subcaptionblock}{114pt}%
    \centering%
    \includegraphics[page=2]{hardness/models}%
    \caption{}%
    \label{fig:models-sst}%
  \end{subcaptionblock}%
  \hfill%
  \begin{subcaptionblock}{114pt}%
    \centering%
    \includegraphics[page=3]{hardness/models}%
    \caption{}%
    \label{fig:models-ft-via-rr}%
  \end{subcaptionblock}%
  \hfill%
  \begin{subcaptionblock}{124pt}%
    \centering%
    \includegraphics[page=4]{hardness/models}%
    \caption{}%
    \label{fig:models-nm}%
  \end{subcaptionblock}%
  \hfill%
  \begin{subcaptionblock}{114pt}%
    \centering%
    \includegraphics[page=5]{hardness/models}%
    \caption{}%
    \label{fig:models-ft-via-nm}%
  \end{subcaptionblock}%
  \hfill%
  \begin{subcaptionblock}{114pt}%
    \centering%
    \includegraphics[page=6]{hardness/models}%
    \caption{}%
    \label{fig:models-ft-via-sst}%
  \end{subcaptionblock}%
  \caption{%
    Models can be defined by the way they are simulated by other models, such as
    \capref{fig:models-rr}~$\RR$ by $\IC$,
    \capref{fig:models-sst}~$\SSt$ by $\IC$,
    \capref{fig:models-ft-via-rr}~$\FT$ by $\RR$,
    \capref{fig:models-nm}~$\NM$ by $\RR$,
    \capref{fig:models-ft-via-nm}~$\FT$ by $\NM$, and
    \capref{fig:models-ft-via-sst}~$\FT$ by $\SSt$.
    Particles are dark blue with white labels, anti-particles light red with black labels.
    Compare~\capref{fig:models-ft-via-nm} and~\capref{fig:models-ft-via-sst} for an illustration of the duality of $\SSt$ and $\NM$.
  }%
  \label{fig:models}%
\end{figure}

Problems concerning $\IC$ may be interesting in themselves (e.g., generalizations of the 15-puzzle~\cite{fifteen-puzzle}), but we are motivated by applications characterized by a lack of precise control.
$\IC$ is useful as a base model allowing us to represent atomic moves of restricted models as sequences of $\IC$~moves.
When the distance particles travel cannot be finely adjusted but individual control signals exist, the setting becomes reminiscent of \emph{Randolph's robots}~\cite{randolph-endm06,randolph-jip17}, leading to the model $\RR$ that recursively calls a move~$\delta^{\IC}_{(p,d)}$ on the neighbor of~$p$ in direction~$d$ until the neighbor is occupied or out of bounds; see Figure~\ref{fig:models-rr}.
Conversely, if distances can be regulated precisely but there is no individual control, one can represent this by the application of $\IC$ moves on all occupied nodes, in the opposite order to the direction of movement (e.g., leftmost nodes first for a move in direction~\ldir{}.)
We recover the single step model this way, see Figure~\ref{fig:models-sst}.
Combining both restrictions results in the full tilt model, see Figure~\ref{fig:models-ft-via-rr}.
Thus, by definition, we have $\delta^{\IC} \succeq \delta^{\RR} \succeq \delta^{\FT}$ and $\delta^{\IC} \succeq \delta^{\SSt}$.

Although~\cite{tilt-fill-icalp} only considered unlabeled configurations, several results transfer to our generalized setting, such as $\SSt \succeq \FT$.
More importantly, $\FT$ is dual to itself, since moving all particles maximally in one direction is the same as moving all anti-particles maximally in the opposite direction.
This suffices to determine the complexity of \vacancy{} in~$\FT$.

\begin{corollary}
\vacancy{} is \PSPACE-complete for simple environments in~$\FT$.
\end{corollary}
\begin{proof}
It is easy to see how a non-deterministic Turing machine can decide \vacancy{} in~$\FT$ using polynomial space, which implies containment in \PSPACE{} by Savitch's theorem~\cite{savitch}.
\PSPACE-hardness follows from Corollary~\ref{cor:hardness-occupancy-ft} and Observation~\ref{obs:occupancy-vacancy-dual} via duality of~$\FT$ to~itself.
\end{proof}

\paragraph*{The dual of the single step model}

We have seen how $\FT$ can be defined in terms of $\RR$ by moving all particles in the order opposite to the direction of movement.
One may as well choose to use the \emph{same} direction for the order of particles as for the direction of movement (e.g., the rightmost particles first for an \ldir{}-move).
This leads to the final model we will consider, illustrated in Figure~\ref{fig:models-nm}.
We refer to this model as the \newterm{Newton~model}~($\NM$) because it gives the impression of particles colliding and transferring their momentum in a way reminiscent of Newton's cradle.

It is fair to wonder at this point what all these models are good for, when we initially set out to tackle the complexity of \vacancy{} in~$\SSt$.
We will next show that $\NM$ and $\SSt$ are closely connected.
Afterwards, we prove limitations of $\RR$, which extend to $\NM$ because $\delta^{\RR} \succeq \delta^{\NM}$.
Combining these arguments with a slight modification to the reduction from Section~\ref{sec:complexity-simple} will allow us to prove \vacancy{} \PSPACE-complete in~$\SSt$.

\begin{lemma}\label{lem:sst-nm-dual}
The single step model is dual to the Newton model, i.e., $\overline{\delta^{\SSt}} = \delta^{\NM}$.
\end{lemma}
\begin{proof}
We only show $\overline{\delta^{\SSt}_{\ldir}} = \delta^{\NM}_{\rdir}$ -- the argument is analogous for other pairs of opposite directions.
Consider an environment~$B=(V,E)$ with a configuration~$C$ and let $p_1, \ldots, p_k \in V$ be a sequence of all occupied nodes ordered from left to right (breaking ties arbitrarily).
Then $w \coloneq (p_1,\ldir) \ldots (p_k, \ldir)$ is a sequence of $\IC$ moves that simulates an \ldir{}-move of $\SSt$ on~$C$, i.e., $\delta^{\SSt}_{\ldir}(C) = \delta^{\IC}_{w}(C)$.
Since moves $\delta^{\IC}_{(p,\ldir)}$ and $\delta^{\IC}_{(q,\ldir)}$ commute if $p$ and~$q$ are not in the same row segment, we can re-order the $p_i$ such that nodes from the same row segment appear in contiguous sub-sequences.
Consider a row segment~$S \subseteq V$ with nodes $s_1, \ldots, s_{|S|}$ ordered from left to right, and let $w_S$ be the full sub-sequence of~$w$ affecting nodes from~$S$.
Since $\delta^{\IC}_{(p,\ldir)}$ is the identity if $p$ is the leftmost node in~$S$ or contains an anti-particle, we can choose $w'_S \coloneq (s_2, \ldir) \ldots (s_{|S|},\ldir)$ such that $\delta^{\IC}_{w'_S}(C) = \delta^{\IC}_{w_S}(C)$.
Observe that the dual move~$\delta^{\IC}_{(p,\ldir)}$ for a node~$p$ with left neighbor~$u$ is~$\delta^{\IC}_{(u,\rdir)}$.
Thus, $\overline{\delta^{\IC}_{w'_S}} = \delta^{\IC}_{w''_S}$ for $w''_S \coloneq (s_1,\rdir) \ldots (s_{|S|-1},\rdir)$.
Apart from a possible prefix of identity moves, $w''_S$ is precisely the simulation of an $\rdir$\nobreakdash-move of~$\NM$ by~$\IC$ on the complement of~$S$.
Applying the same argument to all row segments and concatenating the resulting sub-sequences to form~$w'$ from all $w'_S$ and $w''$ from all $w''_S$, we attain $\delta^{\SSt}_{\ldir}(C) = \delta^{\IC}_{w'}(C) = \texttt{comp}(\delta^{\IC}_{w''}(\texttt{comp}(C))) = \texttt{comp}(\delta^{\NM}_{\rdir}(\texttt{comp}(C)))$, i.e., $\overline{\delta^{\SSt}_{\ldir}(C)} = \delta^{\NM}_{\rdir}(C)$.
\end{proof}

Figures~\ref{fig:models-ft-via-nm}--\subref{fig:models-ft-via-sst} illustrate the duality of~$\SSt$ and~$\NM$ in the context of simulating~$\FT$.

\begin{figure}[tb]
  \centering%
  \begin{subcaptionblock}{110pt}%
    \centering%
    \includegraphics[page=1]{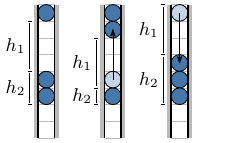}%
    \caption{}%
    \label{fig:rr-dead-end-a}%
  \end{subcaptionblock}%
  \hfill%
  \begin{subcaptionblock}{110pt}%
    \centering%
    \includegraphics[page=2]{hardness/rr-dead-end}%
    \caption{}%
    \label{fig:rr-dead-end-b}%
  \end{subcaptionblock}%
  \hfill%
  \begin{subcaptionblock}{110pt}%
    \centering%
    \includegraphics[page=3]{hardness/rr-dead-end}%
    \caption{}%
    \label{fig:rr-dead-end-c}%
  \end{subcaptionblock}%
  \caption{%
    Three cases from the proof of Lemma~\ref{lem:dead-end-rr}.
    \capref{fig:rr-dead-end-a}~$h_2 \geq 2$.
    \capref{fig:rr-dead-end-b}~$h_2 = 1$.
    \capref{fig:rr-dead-end-c}~$h_2 = 0$.
  }%
  \label{fig:rr-dead-end}%
\end{figure}

\begin{lemma}\label{lem:dead-end-rr}
Let $B=(V,E)$ be an environment, $S \subseteq V$ a column segment with a dead end of length~$2 \ell$ at its top, for some $\ell > 0$, $p$ the $\ell$th node from the top in~$S$, and $C$ a configuration of~$B$ with no occupied node in~$S$.
If there is no sequence $w \in (V \times D)^*$ such that $\ell$~nodes of~$S$ are occupied in~$\delta^{\RR}_w(C)$, then no such sequence exists that leads to $p$ being occupied.
\end{lemma}
\begin{proof}
Assume that at most $\ell - 1$ nodes in~$S$ can be occupied at once, no matter the sequence of $\RR$-moves applied to~$C$.
Let $S' \subseteq S$ be the dead end of length~$2\ell$ at the top of~$S$.
After every move, we partition the nodes of~$S'$ into maximally contiguous blocks of occupied and free nodes.
Let~$h_1$ be the size of the first block of free nodes in~$S'$ in top-to-bottom order, $h_2$ the size of the occupied block below ($h_2 = 0$ if no such block exists), and $k_S$ the number of occupied nodes in~$S'$.

We now show that $h_1 \geq \ell + 1$ holds after the application of any sequence of moves.
Initially, we have $h_1 = 2\ell$.
To prove that $h_1 \geq \ell + 1$ is preserved by every possible move, we consider three cases.
\begin{description}
\item[$h_2 \geq 2$] The only moves that affect the first free block are \udir{} on the particle below and \ddir{} on the particle above (if one exists).
  Neither move changes~$h_1$, see Figure~\ref{fig:rr-dead-end-a}.
\item[$h_2 = 1$] As in the previous case, a \ddir{}-move on the particle above does not change~$h_1$.
  \udir{} or \ddir{} on the single particle constituting the occupied block below may increase~$h_1$ but cannot possibly decrease it, see Figure~\ref{fig:rr-dead-end-b}.
\item[$h_2 = 0$] Again, moving a particle above down cannot increase $h_1$ and may even decrease it, see Figure~\ref{fig:rr-dead-end-c}.
  The only way to increase~$h_1$ is a \udir{}-move on a particle in $S \setminus S'$ when $h_2 = 0$, which preserves $h_2 = 0$.
  However, $h_2 = 0$ implies $h_1 \geq 2\ell - k_S$.
  Since we assumed $k_S < \ell$, we retain $h_1 \geq \ell + 1$.
\end{description}
Thus, the first free block covers at least $\ell + 1$ contiguous nodes of~$S'$, which includes~$p$.
\end{proof}

\restateVacancyHardSSt*
\begin{proof}
Since a non-deterministic Turing machine can clearly apply arbitrary sequences of $\SSt$-moves in polynomial space, the problem is contained in $\NPSPACE=\PSPACE$~\cite{savitch}.

Observation~\ref{obs:occupancy-vacancy-dual} and Lemma~\ref{lem:sst-nm-dual} imply that \PSPACE-hardness of \vacancy{} in~$\SSt$ follows from \PSPACE-hardness of \occupancy{} in~$\NM$.
Recall the reduction from Theorem~\ref{thm:hardness-main-thm}.
That proof itself requires no changes to apply to $\NM$, because $\FT$ and~$\NM$ behave identically when no two particles are in the same segment, and once a particle enters a trap, it is also true in~$\NM$ that from then on there will always be a particle in the trap.
Thus, it is \PSPACE-hard to decide in $\NM$ if all $m+1$ particles can be moved to the leftmost column.

To prove hardness of \occupancy{} in $\NM$, we use a slightly different modification than the one we used for~$\FT$.
Instead of adding two dead ends of length~$m+1$ to both ends of the leftmost column, we add a single dead end of length~$2m+2$ to its top, and ask if it is possible to occupy the node~$p$ at position $m+1$ from the top.

Assume there is a sequence of $\NM$-moves that moves all $m+1$ particles to the leftmost column.
Then, appending $\udir^{m+1}$ leads to $p$ being occupied.

Conversely, assume there is no sequence of $\NM$-moves that moves all $m+1$ particles to the leftmost column.
Then, there is no sequence that leads to $p$ being occupied, due to Lemma~\ref{lem:dead-end-rr} and the fact that $\delta^{\RR} \succeq \delta^{\NM}$.
\end{proof}